%% file: main.tex
\documentclass{egpubl}
\usepackage{eurovis2025}

\SpecialIssuePaper         

\CGFccby

\usepackage[T1]{fontenc}

\usepackage{cite}  
\BibtexOrBiblatex
\PrintedOrElectronic
\ifpdf \usepackage[pdftex]{graphicx} \pdfcompresslevel=9
\else \usepackage[dvips]{graphicx} \fi

\usepackage{egweblnk}
\usepackage{multirow}
\usepackage{times}
\usepackage{latexsym}
\usepackage{booktabs}
\usepackage{amsmath}
\usepackage{enumitem}
\usepackage{scalefnt}
\usepackage{float}
\usepackage{xspace}
\usepackage{wrapfig}
\usepackage{soul}
\usepackage{subcaption}
\usepackage[table]{xcolor}
\usepackage{array}
\usepackage{makecell}
    \newcolumntype{P}[1]{>{\centering\arraybackslash}p{#1}}
    \newcolumntype{M}[1]{>{\centering\arraybackslash}m{#1}}

\newcommand{\hlc}[2][yellow]{{%
    \colorlet{foo}{#1}%
    \sethlcolor{foo}\hl{#2}}%
}

\usepackage{microtype}
\usepackage{csquotes}

\newcommand{\etal}{\textit{et al.}\xspace}

\title{MatplotAlt: A Python Library for Adding Alt Text to Matplotlib Figures in Computational Notebooks}

\author[K. Nylund \& J. Mankoff \& V. Potluri]
{\parbox{\textwidth}{\centering Kai Nylund$^{1}$ \orcid{0009-0006-3956-7890}
        and Jennifer Mankoff$^{1}$ \orcid{0000-0001-9235-5324}
        and Venkatesh Potluri$^{2}$ \orcid{0000-0002-5027-8831}
        }
        \\
{\parbox{\textwidth}{\centering $^1$Paul G. Allen School of Computer Science \& Engineering, University of Washington \\
    $^2$University of Michigan School of Information\\
	{\texttt{knylund@cs.washington.edu}}
       }
}
}

\begin{document}
    \input{./figures/teaser}
    \maketitle
    \begin{abstract}
      \input{./content/abstract}
    \end{abstract}
    
    \input{./content/introduction}
    \input{./content/related_work}
    \input{./content/system}
    \input{./content/datasets}

\input{./content/evaluation}
    \input{./content/discussion}
    \input{./content/acknowledgements}
    \newpage
    \bibliographystyle{eg-alpha-doi} 
    \bibliography{main}  
    \newpage

    \appendix
    \input{./supp_content/example_gallery}
    \input{./supp_content/templates}
    \input{./supp_content/vlm_prompts}
    \input{./supp_content/ngram_similarity}
    \input{./supp_content/shuffling}
    \input{./supp_content/correlations}
\end{document}

%% file: figures/teaser.tex
\teaser{
    \includegraphics[alt={A screenshot of a Jupyter notebook with annotations. At the top is a Python code cell with the annotation "Matplotlib code to generate a barplot". Below the Matplotlib code is the line show\_with\_alt(desc\_level=3, methods=["markdown", "img\_file"]). An annotation indicates this line is "calling MatplotAlt's show\_with\_alt function to generate and surface alt text using the 'markdown' and 'img\_file' methods". Below the code cell is alt text for the generated bar plot in markdown, with the annotation "Heuristic-based alt text is displayed in markdown output". Finally, at the bottom of the figure is the generated barplot displaying average yearly hours of sunshine for each month. The last annotation indicates that this "Matplotlib figure is displayed and saved as an image file with embedded alt text".}, width={\linewidth}]{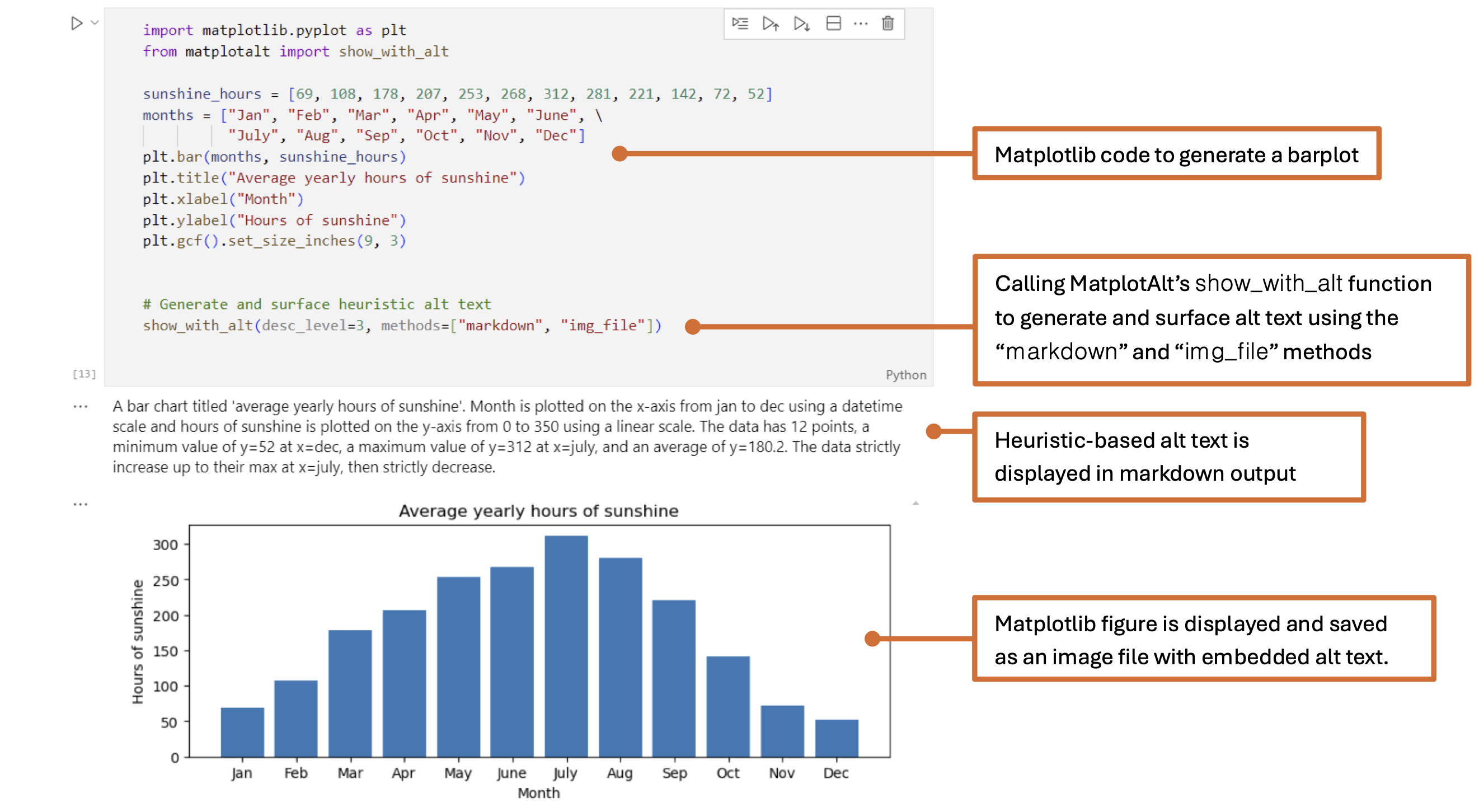}
    \centering
    \caption{\textbf{MatplotAlt can be used to generate and surface Matplotlib alt text in a single line of code.} In this example, we call MatplotAlt's \texttt{show\_with\_alt} function after creating a Matplotlib bar chart to display heuristic-based alt text in markdown and embedded in a saved version of the figure. \texttt{desc\_level=3} indicates that the description includes encodings, statistics, and trends. MatplotAlt also provides options to embed alt text directly in Jupyter figures, and generate descriptions using vision language models.}
    ~\\
    \label{fig:teaser}
}

%% file: content/abstract.tex
We present MatplotAlt, an open-source Python package for easily adding alternative text to Matplotlib figures. MatplotAlt equips Jupyter notebook authors to automatically generate and surface chart descriptions with a single line of code or command, and supports a range of options that allow users to customize the generation and display of captions based on their preferences and accessibility needs.
Our evaluation indicates that MatplotAlt's heuristic and LLM-based methods to generate alt text can create accurate long-form descriptions of both simple univariate and complex Matplotlib figures. We find that state-of-the-art LLMs still struggle with factual errors when describing charts, and improve the accuracy of our descriptions by prompting GPT4-turbo with heuristic-based alt text or data tables parsed from the Matplotlib figure.

%% file: content/introduction.tex
\section{Introduction}

Computational notebooks centralize code, natural language, and visualizations in a single interactive medium. Because of their versatility and ease of use, Jupyter notebooks have become popular for teaching, communicating results and conducting analysis. 
Despite their ubiquity, notebooks remain inaccessible to blind and visually impaired (BVI) users due to the authoring practices, tools, and infrastructures used in their creation and sharing~\cite{potluri2023notably}. Notably, in their analysis of 100,000 notebooks, Potluri and Singanamalla \etal{} found that 99.81\% of programmatically generated images did not have associated alternative  text~\cite{potluri2023notably}, which is a critical accessibility barrier \cite{elavsky2022accessible, world2008web}.

From a sample of 10 million notebooks, the JetBrains Datalore team found that about 42\% had programatically generated images \cite{Guzharina_undated-ai}.
The vast majority of these were found to be data visualizations created with the Matplotlib or Seaborn libraries, neither of which contain methods to easily embed image descriptions or alternative text~\cite{potluri2023notably,MatplotlibAccessibility2019}. Other visualization tools like Altair 
\cite{VanderPlas2018} are beginning to include options to create accessible text structures for screen reader users, but these updates are not compatible with most inaccessible notebooks on the web that do not use these libraries \cite{Guzharina_undated-ai}. Additionally, no popular computational notebook software currently supports the addition of descriptive text to displayed images.

To help notebook users create and consume image descriptions for the majority of notebooks, we present \emph{MatplotAlt}, a Python package to add alt text to Matplotlib figures. MatplotAlt provides functions to automatically generate descriptions from a Matplotlib figure object, and several methods to embed and export its alt text.
These automatically generated captions are informed by Lundgard and Satyanarayan's four-level model of semantic content \cite{lundgard2021accessible}, a set of guidelines to effectively describe data visualizations. To create both accurate and fluent chart descriptions, MatplotAlt provides two options for generating alt text: a heuristic based approach which directly uses figure attributes and data, and a vision-language model (VLM) based method which takes the full image as input. 
While state-of-the-art VLMs can generate relevant alt text \cite{chen2023sharegpt4v, zhang2024good, singh2024figura11y}, we confirm that they are still prone to factual errors on both simple univariate and complex Matplotlib figures. We show that prompting VLMs with template-based alt text is a simple way to increase their accuracy and similarity to human-written descriptions. In this work, we contribute:

\begin{itemize}
\item MatplotAlt, the first Python library to support programmatic generation, inclusion, and dissemination of customizable alternative texts for Matplotlib images in computational notebooks.
\item Quantitative and qualitative evaluations of MatplotAlt's methods for generating alt text, including a detailed analysis of errors in VLM descriptions.
\item Strategies to improve the quality of VLM descriptions for Matplotlib figures.
\end{itemize}
MatplotAlt's integration into the notebook pipeline, ease of use, and readiness for VLM-generated descriptions have the potential to make notebooks accessible to BVI users at scale and serve as a blueprint for plotting libraries to build in native support for alt text. The scenarios below show how MatplotAlt can help users perceive Matplotlib images and work with notebooks. We make Matplotalt and our evaluation datasets publicly available at \URL{https://github.com/make4all/matplotalt}.

\input{content/example_scenario}

%% file: content/example_scenario.tex

\subsection{Example Scenarios}
\label{sec:example_scenarios}

Aida is a BVI developer in a data analysis team comprised of BVI and sighted scientists. Aida wants to visualize data from their research project using Matplotlib in a Jupyter notebook, then add alt text to their figures for future screen reader users and their own reference. To include automatically generated alternative text in the HTML output of the notebook, they replace calls to \texttt{matplotlib.pyplot.show} with MatplotAlt's \texttt{show\_with\_alt} function. Rory, Aida's sighted colleague adds additional analysis that results in complex figures. Rory uses \texttt{generate\_alt\_text}, which returns generated alt text for the last figure as a string. They check for factual correctness and manually add their own context and insights to the returned string. Finally, Rory displays their customized alt text with \texttt{add\_alt\_text} for Aida to review.

Dez is a BVI student trying to learn a new Python package by reading its documentation page. Using a screen reader, they notice most of the examples are in a computational notebook with several Matplotlib figures. Unfortunately, the only text read when they select one of the images is ``\emph{No description has been provided for this image}'', which is the default in Jupyter. Dez downloads the notebook and runs MatplotAlt's \texttt{alttextify} command from the terminal to automatically create and embed descriptions for each figure. After reading the generated alt text, Dez feels several figures are still lacking context, so they rerun \texttt{alttextify} with a VLM by passing an API key.  Once Dez has read both versions of figure alt text they have a clearer idea of how to use the library in their coursework. Satisfied with this output, Dez reruns \texttt{alttextify} with the \texttt{-s new\_cell} argument to store generated alt text as a code comment. Dez converts this updated notebook into a python file using standard nbconvert tools, then continues to work on the assignments with the alt-text and starter code giving them the necessary information to use the library effectively.

These scenarios describe how the various options provided by MatplotAlt equip a BVI expert to exercise agency by generating customized alternative texts, and overcome accessibility barriers previously posed by poor infrastructure and authoring practices.

%% file: content/related_work.tex
\begin{figure*}[t!]
    \centering
    \includegraphics[alt={A block diagram depicting the MatplotAlt system. There are boxes for a Matplotlib figure, chart type classifier, the generate\_alt\_text function, the generate\_api\_alt\_text function, and the add\_alt\_text function. Two arrows from the matplotlib figure block indicate that "Figure features and data" are passed to generate\_alt\_text, and the image is passed to generate\_api\_alt\_text. The chart type classifier box takes the figure features and data and outputs type to the generate\_alt\_text function. The generate\_alt\_text block contains the text "params: semantic level, stats, trends, include table, other options", and the generate\_api\_alt\_text box contains "params: API key, model prompt, include heuristic alt or table, max tokens". An arrow indicates that heuristic alt text and data tables are optionally passed from generate\_alt\_text to generate\_api\_alt\_text for use in model prompts. Finally, arrows show that heuristic alt text and VLM alt text are passed from generate\_alt\_text and generate\_api\_alt\_text respectively to the add\_alt\_text function, which contains the text "methods: Embed in HTML, Output as markdown or new cell, Export as image or text file"},width=\textwidth]{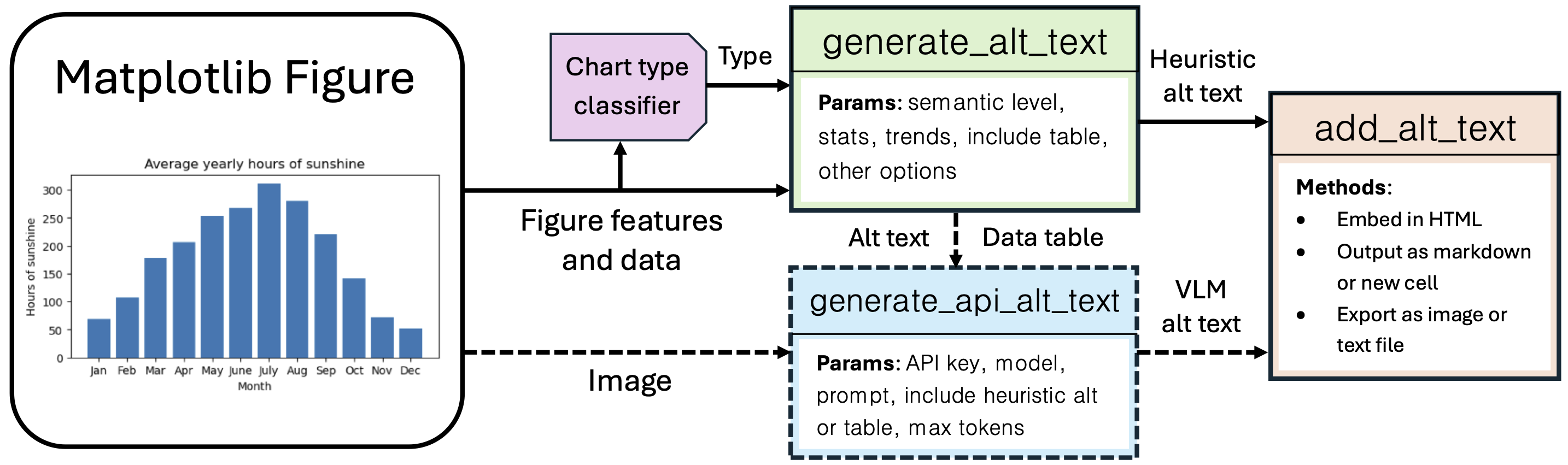}
    \caption{\textbf{The MatplotAlt system.} Heuristic-based alt text of a given semantic level is generated from Matplotlib figure attributes using the \texttt{generate\_alt\_text} function. Alternatively, VLM-based descriptions can be generated with the figure as input using \texttt{generate\_api\_alt\_text}. Descriptions are then embedded in the notebook or exported using \texttt{add\_alt\_text}.}
    \label{fig:MatplotAlt-architecture}
\end{figure*}

\section{Related Work}
\label{sec:related-work}
MatplotAlt is motivated by a growing set of literature on generating descriptive alt text and the (in)accessibility of visualization systems. We provide a brief summary about accessible visualizations, motivate the need to make them accessible in the context of computational notebooks, and provide relevant background on emerging automatic description techniques.

\subsection{Accessibility of Web Visualizations}

Despite improvements in systems to generate and publish alt text, most figures on the web remain inaccessible to screen reader users. Extending the Web Content Accessibility Guidelines (WCAG), Elavsky \etal{} developed Chartability, a set of heuristics to evaluate the accessibility of data visualizations, including for keyboard navigation, screen reader inspection, and cognitive barriers \cite{elavsky2022accessible}. Their work highlights the changing nature of accessibility standards, and the need for context-specific guidelines in visualization.

Visualizations on the web rarely adhere to these guidelines or contain even basic accessibility support necessary for BVI users to gain insights from data.
BVI users report several common pain points, including invisible or incomprehensible figures, a lack of description of trends and axes, and the inability to expose data in tables and arrays~\cite{sharif2021understanding}. 
Visualization dashboards, which are often presented as web interfaces, also frequently contain barriers like inconsistent semantic structure and unsurfaced changes \cite{srinivasan2023azimuth}. 

These barriers are compounded by a lack of practical methods for creating accessible outputs. Joyner \etal{} survey visualization practitioners and identify complex / interactive figures and lack of accessibility support in visualization tools as obstacles preventing the creation of accessible figures \cite{joyner2022visualization}. Several systems aim to make creating alt text easier \cite{mack2021designing, singh2024figura11y, moured2024alt4blind}.
Mack \etal{} \cite{mack2021designing} explore using a template interface for authoring figure descriptions in PowerPoint with separate boxes to describe subjects, interactions, and other features. Singh \etal{} support authoring alt text for scientific publications with an interface that extracts figures, captions, and data from uploaded PDFs, then provides appropriate guidelines and suggestions generated by a large language model (LLM) \cite{singh2024figura11y}. They find that most users prefer the extra assistance provided by templates, and that they help users know what to include in alt text \cite{mack2021designing}. Singh \etal{} also find that LLM-generated alt texts were a useful starting point for most users \cite{singh2024figura11y}. These results inform our decision to support automatically generating starter alt text using either a template (with no user input) or VLM. MatplotAlt extends the capabilities of this growing body of work to Python figures in notebooks, allowing us to more easily extract data and features from programmatically generated charts, and address the specific accessibility barriers of this environment.

Guidelines for describing figures have also improved in the last three years. Lundgard and Satyanarayan define a theory-grounded four-level model of semantic content for describing data visualizations \cite{lundgard2021accessible}. Different levels include information about encodings and axes (level 1), relevant statistics (level 2), trends (level 3), and broader context about the chart (level 4). MatplotAlt uses these guidelines to generate descriptions depending on users' desired level of detail. 

\subsection{Automatically Generating Alt Text}
Several systems focus on automatically generating accessible figures from data using heuristics and templates. Mirri \etal{} explore using CSV content to create screen reader compatible XML graphics including basic alt information \cite{mirri2017towards}, and Sharif \etal{} develop a jQuery plugin which summarizes HTML/Json figure types, values, and basic statistics \cite{sharif2018evographs}. One limitation of alt text created from static templates alone is their lack of interactivity \cite{sharif2021understanding}. Allowing a range of exploration methods, BrailleR \cite{brailler2024} supports both creating templated textual descriptions of graphs in R and exploring different chart features with a screen reader in an interactive SVG. MatplotAlt aims to similarly support exploring charts in Python notebooks with both templated and VLM descriptions.

Recently, there has been growing interest in generating captions and interactive descriptions for data visualizations using vision transformers. MatCha \cite{liu2022matcha} and UniChart \cite{masry2023unichart}, for instance, are both pretrained image-to-text Transformer \cite{vaswani2017attention} models to automatically summarize and respond to user questions about figures. ChartVLM \cite{xia2024chartx} improve on chart QA tasks by scaling up their model and using a classifier to transform natural language instructions into a discrete set of tasks.

Though these efforts demonstrate the viability of vision transformer models to generate chart captions and support conversational data analysis, they do not account for the nuanced information needs of BVI experts. To address this gap, Tang \etal{} curate VisTex \cite{tang2023vistext}, a dataset of image / caption pairs adhering to the four-level semantic model, which is informed by preferences and needs of BVI users. VLMs are also prone to errors that can mislead users relying on these descriptions \cite{zhang2024good, kantharaj2022chart}. Tang \etal{} qualitatively analyze errors produced by their models and find that approximately half of generated statements were factually incorrect \cite{tang2023vistext}. Using LLMs to generate alt text can also create new barriers like API costs, large downloads, and knowledge of natural language processing libraries. 

At the same time, heuristic-based alt text is prone to repetition over multiple charts and may be unable to capture context that is essential to interpreting figures. These limitations motivate our decision to provide options to generate both heuristic and VLM-based alt text. For high-stakes analysis in the medical domain, for example, MatplotAlt's dry but interpretable templated descriptions are likely more appropriate \cite{rudin2019stop} compared to those generated by a VLM that potentially contain errors and hallucinations. Descriptions for personal learning on well-documented problems, however, may benefit from the adaptability and extra context provided by VLMs \cite{rasul2023role}. 

\subsection{Accessibility of Computational Notebooks}
Despite their ubiquity, computational notebooks remain inaccessible to BVI users. Computational notebook usage is growing, with the number of .ipynb files on Github increasing from 200,000 in 2015 to 2.5 million in 2018 \cite{perkel2018jupyter}, and a dataset of 10 million Jupyter notebooks released in 2020 \cite{Guzharina_undated-ai}. Several authors have even suggested making notebooks the primary artifact of scientific publications \cite{somers2018scientific, rowe2020potential, caprarelli2023notebooks}; but fundamental accessibility barriers remain unaddressed. Building on prior work documenting pain points in computational notebooks \cite{chattopadhyay2020s}, Potluri \etal{} analyze the inaccessibility of Jupyter for BVI developers and users \cite{potluri2023notably}. In their evaluation of 100,000 notebooks, they found that 99.81\% of programmatically generated images, most of which were created with Matplotlib or seaborn, lacked associated alt text. They also identify a range of common WCAG failures in notebooks including low color contrast, incorrectly nested headers, and a lack of correctly formatted data tables. 

Based on these errors, the authors provide a set of guidelines for improving the accessibility of Jupyter notebooks. These include using the four-level semantic model to generate and include alt text for plotting libraries like Matplotlib using the figure object, and automatically generating markdown tables for applicable visualizations. MatplotAlt is directly inspired by these recommendations, with the goal of providing an easy interface to include alt text and data tables for figures in Jupyter notebooks. 


%% file: content/system.tex
\section{The MatplotAlt System}
\label{sec:the_MatplotAlt_system}
We describe the algorithms, templates, and libraries we use to automatically generate and surface alt text for Matplotlib figures. Figure \ref{fig:MatplotAlt-architecture} depicts a simplified view of the MatplotAlt system highlighting components responsible for parsing chart features (\S\Ref{sec:chart_type_cls}), generating heuristic (\S\ref{sec:heuristic_alt}) and VLM-based (\S\ref{sec:vlm_alt}) alt text, and surfacing descriptions for screen readers (\S\ref{sec:add_alt_text}). To ensure that accessibility barriers posed by notebook software do not hinder the utility of MatplotAlt, we also provide options to use these tools outside the notebook environment (\S\ref{sec:alttextify}).

\subsection{Inferring Chart Data and Features}
\label{sec:chart_type_cls}
Chart descriptions should contain information about the chart type, and other visual encodings to adhere to the four-level model of semantic content \cite{lundgard2021accessible}. To generate alt text that includes these details, we parse most features and data directly from Matplotlib figures. To get a list of the tick labels on the x-axis, for example, we call Matplotlib's \texttt{get\_xticklabels()} method on the most recently plotted figure. Chart type is one notable exception, as Matplotlib only stores abstract elements like rectangles, lines, and points. MatplotAlt infers type based on these components. If a plot has multiple point objects without any lines, for instance, we label it as a scatterplot. If the figure contains both lines and points, then we infer that it is a line plot. And if the plot contains a quadmesh, it is likely a heatmap or image. In addition to chart features, we also consider how data is internally represented when determining type. Heatmaps and images, for example, contain 2d arrays of values while line plots have an iterable of line objects and values.

One downside of this approach is that it assumes all charts have a single type. For example, our current system cannot correctly classify overlaid bar and line plots in the same figure. Our system also currently fails on interactive and dynamically updated charts. Unlike more complex classification models, however, MatplotAlt can be easily extended to new chart types by adding checks for new attributes without the need for retraining. Extending previous work \cite{kantharaj2022chart, tang2023vistext} that focus on only the one to three most popular chart types, MatplotAlt currently supports ten types of figures: \emph{line, bar, scatter, radial line, pie, strip, contour, heatmap, image}, and \emph{boxplot}. 

\subsection{Generating Heuristic-based alt text}
\label{sec:heuristic_alt}

To automatically generate alt text for Matplotlib figures using templates and heuristics, MatplotAlt provides the \texttt{generate\_alt\_text} function. Like our chart type detection, information used in descriptions is extracted directly from the most recent Matplotlib object.

Users can specify the amount of detail to include in alt text through the \texttt{desc\_level} parameter, based on L1-L3 semantic levels \cite{lundgard2021accessible}:

\begin{itemize}
    \item[\emph{L1:}] Alt text includes the chart's type, title, color encodings, annotations, and the scale and range of each axis.
    \item[\emph{L2:}] Includes L1 plus statistics for each variable. We start with the list of statistics mentioned in the four-level model of semantic content (extrema, outliers, and correlations), and expand to several other easily computed metrics like standard deviation and median. We also adjust the default supported statistics for different chart types. Alt text for scatter plots, for instance, will include minima, maxima, and a line of fit; boxplots' will contain interquartile ranges; and contours' will contain the center point of the min/max contour. These defaults were chosen both based on commonly performed analyses (e.g., line of fit) and feasibility (e.g., boxplots may not display individual points). Users can manually specify which stats to include through the \texttt{stats} parameter.
    \item[\emph{L3:}] Includes L2 plus chart trends such as increasing / decreasing patterns if applicable and the stability (e.g., ``fluctuating'' vs. ``strictly increasing'') of each variable. MatplotAlt currently only supports trends for contiguous two-dimensional data. No L3 description is added to scatter, strip, box, and contour plots. Like, \texttt{stats}, users can specify included trends with the \texttt{trends} parameter.
\end{itemize}

Chart summaries provide an overall picture of Matplotlib figures, but consumers of these artifacts may also want to drill down to the underlying data~\cite{sharif2021understanding,Sharif2023DrilDownExploration}. Elavsky \etal{} suggest the inclusion of tables with figures representing data, an accompaniment overwhelmingly lacking in notebooks \cite{elavsky2022accessible, potluri2023notably}. To support this interactivity, MatplotAlt provides the option to include the underlying chart data as a markdown table in generated alt text. To avoid overly large tables that are hard to navigate with screen readers~\cite{Wang:2022:webDataTables}, we add an adjustable cap on the max number of table rows and columns.
If there are multiple subplots in a figure, heuristic alt text and tables will be generated for each separately. Appendix \S\ref{sec:heuristic_templates} includes more detail about templates used in heuristic-based descriptions. To generate L4 descriptions (including context, explanations, and insights), we explore using a VLM in the next section. 

\subsection{VLM-based alt text}
\label{sec:vlm_alt}

MatplotAlt implements the \texttt{generate\_api\_alt\_text} function to generate figure descriptions with models hosted on OpenAI, Azure, and Huggingface APIs. Like \texttt{generate\_alt\_text}, users can specify the semantic level of VLM alt text with the \texttt{desc\_level} parameter. Prompts to the language model, documented in Appendix \S\ref{sec:vlm_prompts}, include a call to describe figure details based on the given \texttt{desc\_level}, and two example L1-L4 figure descriptions. In addition to L1-L3 alt texts, a \texttt{desc\_level} of 4 can be passed to \texttt{generate\_api\_alt\_text}. Prompts to generate L4 captions will also include the line ``If possible, briefly explain domain-specific insights, current events, and socio-political context that explain the data.'' Unlike heuristic alt text, we note that there is no guarantee VLM descriptions will adhere to semantic guidelines.

To capture both perceptual phenomena (e.g. overlapping points, color patterns) in images and semantic relationships between figure elements, we add chart data to prompts by including either a markdown data table (VLM + table), the heuristic-based alt text for the chart (VLM + heuristic), or both (VLM + table + heuristic). We provide examples of alt text generated using each method in Appendix Table \ref{table:example_descriptions1}, and evaluate each method in \S\ref{sec:evaluation}.

\subsection{Surfacing alt text in Jupyter notebooks}
\label{sec:add_alt_text}

To make figure descriptions visible to screen reader users in Jupyter notebooks, MatplotAlt provides the \texttt{add\_alt\_text} function. Several options are supported through the \texttt{methods} parameter:

\begin{itemize}[leftmargin=*]
    \item ``\texttt{html}'': displays the last figure in html with an alt property containing the given text. This is the default option in MatplotAlt, allowing authors and readers to embed and consume image descriptions without changing the layout of the notebook.
    \item ``\texttt{markdown}'': adds text in markdown to the current cell output. This method visually displays alt text alongside the figure, allowing non-screenreader users to view descriptions. Markdown data tables are surfaced only in cell or saved output.
    \item ``\texttt{new\_cell}'': creates a new (code) cell after this one containing the given text. This method is more disruptive to the notebook structure, but allows users to easily include the description in their code, e.g. as a string variable for further processing.
    \item ``\texttt{img\_file}'': saves the last Matplotlib figure as a png with the given text in its alt property. This allow users to consume only the image and description without having to navigate the inaccessible notebook environment, and enables users to create accessible visualizations for use in other projects.
\end{itemize}


For ease of use, we combine \texttt{generate\_alt\_text} and \texttt{add\_alt\_text} into a single function, \texttt{show\_with\_alt}, to both generate and surface alt text in a single line of code. We do the same for VLM alt text with the \texttt{show\_with\_api\_alt} command. These functions allow users to easily replace calls to \texttt{matplotlib.pyplot.show} with \texttt{show\_with\_alt} to generate accessible figure outputs.

\subsection{Using MatplotAlt outside the notebook environment}
\label{sec:alttextify}

Because the notebook interface itself can be a barrier to BVI users~\cite{JupyterLabAccessibilityStatement2022,JupyterAccessibilityDocs2024}, MatplotAlt also provides the \texttt{alttextify} command to add alt text to all Matplotlib figures in a given notebook. The command, which can be run from PowerShell on Windows or the terminal on Mac, takes a notebook path, and any other arguments supported by \texttt{show\_with\_alt} to generate alt text, then embeds or exports descriptions for each Matplotlib figure in the notebook, allowing BVI users to add consumable alt text to inaccessible notebooks they encounter without ever entering the environment.

%% file: content/datasets.tex
\section{Datasets}
\label{sec:datasets}

\subsection{VisText Captions}

We quantitatively evaluate MatplotAlt's methods for generating L3 alt text on the VisText dataset \cite{tang2023vistext}, which consists of pairs of univariate bar, area, and line charts, and corresponding L1-L3 descriptions written by crowdworkers. VisText also contains table and scenegraph representations of charts, which we use to reverse-engineer matplotlib code for generating each figure. We then call \texttt{generate\_alt\_text} and \texttt{generate\_api\_alt\_text} to create corresponding descriptions for the dev and test splits. These Matplotlib versions of VisText figures differ from the original in a few qualities like color and bar ordering that may slightly reduce our alt text's similarity to the crowdsourced descriptions.

We use GPT4-turbo with default temperature and a max output length of 225 tokens for all L3 VLM-generated alt text. We chose this number as the smallest that did not frequently cutoff captions for the dev set, as prior work has shown that BVI users typically prefer more concise VLM-generated captions \cite{huh2024long}. We evaluate prompting GPT4-turbo to describe charts directly (turbo) and our methods for incorperating chart data and features into prompts (turbo + table, turbo + heuristic, and turbo + table + heuristic).

To match the semantic level of human annotations in VisText, we call all generate methods with \texttt{desc\_level=3}. Because very few of the crowdsourced captions describe chart colors or include statistics other than minima and maxima, we disable descriptions of color encodings and limit stats to the min/max in our heuristic alt text. As a comparison, we reevaluate VisText's pretrained VL-T5 image-only, image + data table, and image + scenegraph models on the original VisText figures. Our datasets are publicly available on MatplotAlt's code repository.

\subsection{Matplotlib Example Gallery Captions}
Although VisText provides a structured set of figures and descriptions, its limitation to univariate bar, area, and line charts makes it unrepresentative of the range of complex visualizations in notebooks on the web. As a more challenging snapshot of Matplotlib figures ``in the wild'', we use notebooks from the Matplotlib gallery (\URL{https://matplotlib.org/stable/gallery}). From 500 notebooks, We call \texttt{alttextify <notebook path> <output> -l 3 -s html img\_file} to generate and export L3 alt text for 200 applicable Matplotlib figures. 

Because many of these charts have data that are not easily represented in scenegraphs or tables, we generate their corresponding descriptions using only L3 heuristic, turbo, and turbo + heuristic methods. While most VisText figures display Statista data for world events (e.g., national debt, infant mortality, company sales), the gallery figures demonstrate the effect of different Matplotlib settings and are therefore tied more closely to the chart format and image itself rather than data.

To summarize, we contribute Matplotlib versions of each of the 12441 VisText figures, a small dataset of 200 figures extracted from Matplotlib Gallery notebooks, and corresponding L3 descriptions for each image using the MatplotAlt generation methods.

%% file: content/evaluation.tex
\section{Evaluation}
\label{sec:evaluation}

We evaluate MatplotAlt's alt text generation methods on the VisText and Matplotlib gallery datasets. We quantitatively measure description length (\S\ref{sec:length_of_alts}), similarity to reference VisText captions (\S\ref{sec:similarity_eval}), and similarity between image and alt text embeddings (\S\ref{sec:blip_descriptiveness}). Next, we qualitatively categorize types of errors in MatplotAlt's descriptions (\S\ref{sec:manual_error_types}) and compare their frequency in different generation methods (\S\ref{sec:manual_error_counts}).

\subsection{Length of Generated Alt Texts}
\label{sec:length_of_alts}

\begin{figure}[!htbp]
    \includegraphics[alt={A figure with two subplots depicting the distribution of alt text lengths for each generation method and dataset. Each subplot contains boxplots for each alt text type with generation method on the vertical y-axis, and length in tokens on the x-axis. The top subplot shows VisText L3 description lengths, with boxplots ordered by mean length including VL-T5 datatable (62.4 tokens), VL-T5 scenegraph (64.9), VL-T5 imageonly (75.0), human (90.0), heuristic (92.8), turbo + heuristic (165.2), turbo (165.5), turbo + table + heuristic (173.1), and turbo + table (174.5). The bottom subplot similarly plots the distibution of alt text lengths on the Matplotlib gallery dataset ordered by mean. These include heuristic (with a mean of 138.6 tokens), turbo + heuristic (176.0), and turbo (184.6). On the gallery dataset, heuristic has a few notable outliers in the 350 to 800 token range.},width=.9\columnwidth]{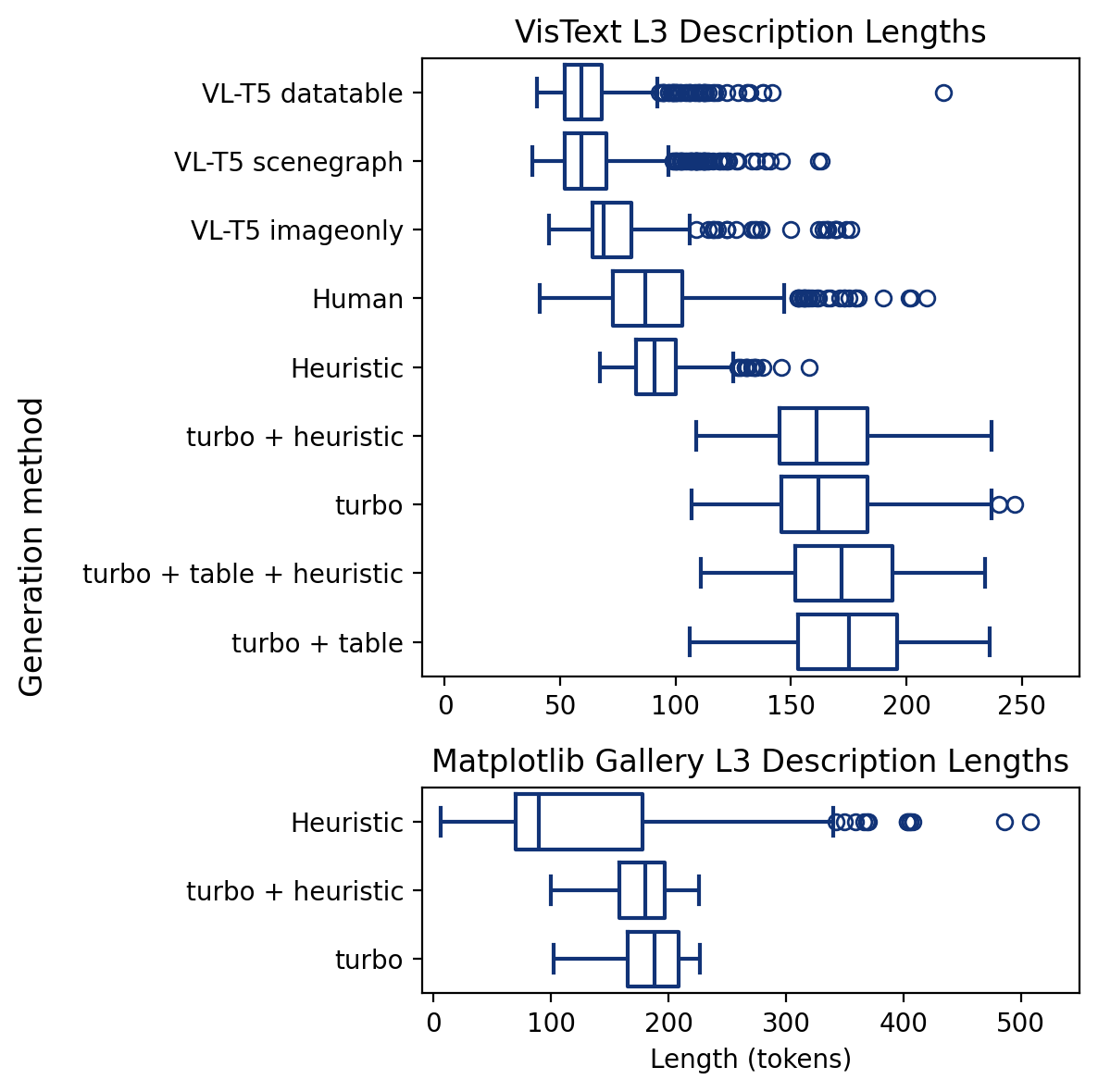}
    \caption{\textbf{MatplotAlt is designed to generate long-form L3 descriptions for Matplotlib figures}. VisText VL-T5 captions were shortest on average, followed by human, heuristic, and GPT4-turbo methods (set to a max output length of 225 tokens). Matplotlib gallery alt texts were generally longer than those for VisText figures.}
    \label{fig:alt_lengths}
\end{figure}

Although we do not know the ideal length of VLM alt texts, prior work shows BVI users may prefer shorter captions because they find the additional information in longer descriptions more repetitive and unnecessary \cite{huh2024long}. In connection to this preference, we characterize the length of VLM alt texts generated with each of our methods. We measure the length of captions using nltk's \texttt{word\_tokenize} function \cite{bird2009natural}. Figure \ref{fig:alt_lengths} shows the distribution of alt text lengths for each generation method and dataset. On VisText, VL-T5 yielded the shortest captions with mean lengths of 62.4, 64.9, and 75.0 for image + datatable, image + scenegraph, and imageonly respectively, followed by human captions (90.0), heuristic (92.8), and the turbo models. We found that VL-T5 captions' length was indicative of a lack of detail, with the model typically only adding one or two sentences about the overall data trend or max values. The crowdsourced captions are similarly concise, but typically contain both trends and descriptions of notable minima and maxima.

Turbo descriptions tended to use most of the token limit, and were more likely to contain additional information about outliers, specific data values, and labels. The extra context from heuristic alt text, data tables, or both increased the length of turbo alt texts even closer to the 225 token limit. For charts with simple trends, however, these longer descriptions were more prone to repetition and unnecessary detail, which are less desired by BVI users \cite{huh2024long}. These results extend to the Matplotlib gallery dataset, with slightly longer heuristic and turbo alt text, likely due to charts with multiple subplots and more features. This difference in lengths between VisText VL-T5 / human and VLM captions indicates our shift to long-form descriptions for Matplotlib figures. The extra information in heuristic and turbo alt texts informs our fine-grained evaluation of error types in \S\ref{sec:manual_error_types} and \S\ref{sec:manual_error_counts}.

\subsection{Similarity to Crowdsourced Captions}
\label{sec:similarity_eval}
Similarity to human captions are widely used to measure the quality of generated alt text for data visualizations~\cite{tang2023vistext,kantharaj2022chart}. In this section, we use two BERT-based metrics to quantify how well MatplotAlt's L3 descriptions match the reference VisText captions.

\subsubsection{Caption Similarity Methods}
For each generation strategy, we measure similarity to the reference human captions using 
Bertscore \cite{zhang2019bertscore}, which averages cosine similarity between the BERT \cite{devlin2019bert} token embeddings for each word (ranging from -1 to 1), and BLEURT \cite{sellam2020bleurt}, which uses a BERT model trained on human ratings of similarity (ranging from slightly below 0 to slightly above 1). If an image has multiple human captions, we take the average of similarities to each. A higher Bertscore indicates that the words in generated captions have meanings similar to the human captions. A cosine similarity of 1, for example, indicates that the sentences are the same, 0 indicates that they are unrelated, and -1 indicates that they have similar but opposite meanings. Similarly, a higher BLEURT indicates that the generated captions more adequately represent the overall meaning of the reference human captions. Bertscore and BLEURT together help measure how semantically similar our generated captions are to the human-written references in VisText.

Because we are not aiming to emulate the style and formatting of the VisText references, Bertscores are likely more effective than exact match n-gram similarity metrics because they capture synonyms. For example, even though ``color'' and ``hue'' are not an exact match, their sentences could still have a high Bertscore due to the high cosine similarity between their word embeddings. BLEURT is similarly less reliant on formatting, and has been shown to have slightly higher agreement with human ratings, but may also be biased towards more fluent text \cite{sellam2020bleurt}.
Although n-gram overlap similarity metrics may not be appropriate due to the formatting and length differences between GPT4 and reference captions, we also measure BLEU \cite{papineni2002bleu}, Rouge \cite{lin2004rouge}, and CHRF \cite{popovic2015chrf} n-gram scores in Appendix \S\ref{sec:ngram_similarity} to compare with prior studies.
\subsubsection{Caption Similarity Results}

\begin{table}
\resizebox{\columnwidth}{!}{%
\begin{tabular}{l | *{4}{M{0.1375\linewidth}}}
\hline
VisText Alt Text Type     & BertScore F1                            & BertScore recall                        & BertScore precision                     & BLEURT                                                \\ \hline
VL-T5 image only          & \cellcolor[HTML]{F6A384}0.8613          & \cellcolor[HTML]{F29173}0.8546          & \cellcolor[HTML]{F7B397}0.8685          & \cellcolor[HTML]{B30326}{\color[HTML]{FFFFFF} -1.042} \\
VL-T5 image + scenegraph  & \cellcolor[HTML]{B4CDFA}\textbf{0.9150} & \cellcolor[HTML]{DEDBDA}0.8939          & \cellcolor[HTML]{80A4FA}\textbf{0.9375} & \cellcolor[HTML]{EFCEBC}-0.2919                       \\
VL-T5 image + table       & \cellcolor[HTML]{B2CBFB}\textbf{0.9161} & \cellcolor[HTML]{DDDCDB}0.8949          & \cellcolor[HTML]{7EA1F9}\textbf{0.9386} & \cellcolor[HTML]{E4D8D1}-0.2653                       \\ \hline
Heuristic                 & \cellcolor[HTML]{ECD1C3}0.8850          & \cellcolor[HTML]{DCDCDD}0.8953          & \cellcolor[HTML]{F4C2AA}0.8753          & \cellcolor[HTML]{A0BFFE}-0.1526                       \\
turbo                     & \cellcolor[HTML]{EFCEBC}0.8826          & \cellcolor[HTML]{DCDCDD}0.8954          & \cellcolor[HTML]{F6B99D}0.8705          & \cellcolor[HTML]{99BAFE}-0.1439                       \\
turbo + table             & \cellcolor[HTML]{EFCEBC}0.8825          & \cellcolor[HTML]{D9DCE0}0.8974          & \cellcolor[HTML]{F7B397}0.8683          & \cellcolor[HTML]{95B7FE}-0.1370                       \\
turbo + heuristic         & \cellcolor[HTML]{E8D5CA}0.8875          & \cellcolor[HTML]{CFD9EA}\textbf{0.9022} & \cellcolor[HTML]{F5BFA5}0.8736          & \cellcolor[HTML]{779AF6}\textbf{-0.0944}              \\
turbo + table + heuristic & \cellcolor[HTML]{E8D5CA}0.8865          & \cellcolor[HTML]{CFD9EA}\textbf{0.9024} & \cellcolor[HTML]{F6BA9F}0.8713          & \cellcolor[HTML]{7B9EF8}\textbf{-0.1000}              \\ \hline
\end{tabular}}
    \caption{\textbf{Adding heuristic alt text to GPT4-prompts can steer generations closer to human-written figure captions.} Similarity between reference and generated descriptions for each method. Higher (blue) is better for each metric. VisText VL-T5 models produce text with the highest BertScore precision and F1, while turbo + heuristic generations typically have the highest Bertscore recall and BLEURT.}
    \label{table:similarity_scores}
\end{table}

Table \ref{table:similarity_scores} shows the similarity scores between Matplotalt and crowdsourced VisText descriptions for each generation strategy. 
We found that adding heuristic alt text in prompts to GPT4-turbo increased similarity to the human-written VisText descriptions across all metrics for both turbo + heuristic and turbo + table + heuristic, and yielded the highest Bertscore and BLEURT among our generation methods. These improvements indicate that prompting with starter alt text is a simple way to steer the style of generated captions closer to ground truth statements. Turbo + table had scores close to turbo, 
with small increases in 
Bertscore, indicating data tables do not significantly change the style of generated captions.

The finetuned VL-T5 models from VisText had higher 
Bertscores, but lower BLEURT. The higher Bertscore is not surprising because the VL-T5 models were trained on pairs of captions and data / images from VisText, while our generation methods are not designed to emulate the prose and length of human captions in the dataset. We find that most of these increases come from the precision component of F1 scores, while turbo descriptions have higher 
Bertscore recalls. This suggests that the GPT4-generated captions contain more of the semantic meaning in human references, and that their lower overall scores may be due more to the additional context and length of VLM descriptions compared to the human written captions.

\subsection{Similarity Between Generated Descriptions and Images}
\label{sec:blip_descriptiveness}
Similarity scores can help quantify how well generated alt texts adhere to the style and content of human descriptions but are still dependent on the length and prose of human writers. Reference-based metrics have also been shown to be biased toward sighted user ratings of alt text over BVI user ratings \cite{kapur-kreiss-2024-reference}.
In an attempt to evaluate Matplotalt's descriptions independently of how well they match the dataset-specific formatting of human alt texts, we measure the similarity between caption and image embeddings.
\subsubsection{Multimodal Similarity Methods}
We use the pretrained BLIP \cite{li2022blip} model to measure how well our generated captions match their corresponding figures. BLIP is a multimodal encoder-decoder vision transformer model \cite{dosovitskiy2020image} pretrained on a variety of tasks including binary classification for whether a caption matches an associated image. Using BLIP, we measure the probability that a description matches its corresponding figure, and the cosine similarity between encoded captions and figures. A high average probability suggests our captions are a good match for their corresponding Matplotlib figures, and a high cosine similarity (ranging from -1 to 1) indicates that the encoded representations of each figure and caption are aligned, suggesting that their content is similar. We average scores for all images and corresponding captions on both the VisText and Matplotlib gallery datasets. Because BLIP was not specifically trained on data visualizations, we assess its relevance for this setting in Appendix \S\ref{sec:shuffling_experiments}.

\subsubsection{Multimodal Similarity Results}
Our BLIP scores on VisText were inconclusive (shown in our Appendix Table \ref{table:shuffled_blip_scores}) with all matching probabilities greater than 0.9975, and all cosine similarities within $0.504 \pm 0.005$. Our results on the Matplotlib gallery dataset show greater variation. GPT4-turbo generations yielded the highest average matching probability (\textbf{0.9766}) and cosine similarity (\textbf{0.4879}) in this setting, followed closely by turbo + heuristic (0.9629 / 0.4717), with a larger gap to the heuristic captions (0.9270 / 0.4263). Excepting the limitations of BLIP, these scores imply that GPT4 alt texts more closely match the challenging Matplotlib gallery figures than the heuristic descriptions, and that adding heuristics in the prompt slightly reduces this alignment. At the same time, the high matching probabilities in VisText and gallery datasets suggest that BLIP alone cannot measure caption \emph{correctness}. Next, we contrast these similarity measures with a qualitative evaluation of caption errors.

\subsection{Types of Errors in Generated Descriptions}
\label{sec:manual_error_types}

We summarize our qualitative observations on the types of errors across generation strategies supported by MatplotAlt. We randomly sample 50 charts from the VisText test and Matplotlib gallery datasets (\S\ref{sec:datasets}) to determine the types of errors that occur in generated descriptions. Including all of our generation methods, each VisText image had 5 descriptions, and each gallery image had 3. The lead researcher annotated a total of 400 captions. Our initial set of error types included those documented in VisText, with the researcher adding new types as they occurred in the descriptions \cite{tang2023vistext}.

To assess relevance of these categories, two researchers on the team (including the primary) discussed the descriptions for 20 randomly selected images from each of the two sets of annotated data: adding, changing, and removing assigned error types as needed. One of the researchers involved in this process used a screen reader full time; the descriptions they annotated would serve as their ground truth in a real-world scenario. We limited this discussion to 20 images from each dataset as both researchers agreed no new error types were being discovered.

From this initial annotation, we identified 13 types of errors. Six of the error types correspond to concrete chart features in descriptions, including:
\begin{itemize}[noitemsep]
    \item \textbf{Value:} error in dependent variable value (e.g. wrong max value).
    \item \textbf{Identity:} error in independent variable (e.g. wrong max location).
    \item \textbf{Chart type:} wrong chart type or orientation (e.g. "line chart" instead of "scatterplot").
    \item \textbf{Axis:} wrong axis range, ticks, or scale.
    \item \textbf{Label:} error in transcribing titles, axis labels, or other figure text.
    \item \textbf{Trend:} wrong direction or stability of trends (e.g. ``strictly increase'' vs. ``generally decrease'').
\end{itemize}
In addition to these factual errors, we identify five common types of formatting mistakes:

\begin{itemize}[noitemsep]
    \item \textbf{Cutoff:} description ends in the middle of a sentence or claim.
    \item \textbf{Missing data context:} description is missing key context about L1-L3 features necessary to understand the chart. This includes failing to mention specific values for statistics and trends, even if the rest of the description is correct.
    \item \textbf{Unnecessary context:} Unnecessary or irrelevant L4 context is included in L3 descriptions.
    \item \textbf{Number name:} Failure to convert number places (e.g. ``10,000 thousand'' instead of ``10 million''), which causes the same denomination like `thousand' to be read twice by a screen reader.
    \item \textbf{Repetition:} The same claim is repeated multiple times.
\end{itemize}
Five of the error types (value, identity, trend, nonsense, and repetition) are also documented in VisText. We combine their direction and stability errors into a single ``trend'' category for simplicity.


Notably, we also distinguish between two types of model hallucinations that do not fit into the other error categories: ``\textbf{nonsense}'', which are identifiable without ground truth information, and ``\textbf{deceptive}'', which are undetectable without perceiving the image. While weaker language model failures are often obvious (e.g. containing disruptive grammar errors or repetition), we included this distinction because GPT4's hallucinations overwhelmingly fit in with the rest of the description, but contain a nonexistent explanation for a trend or confidently describe a chart feature that does not exist. For example, consider the following turbo description for a chart with two line subplots: ``\emph{In the Periodogram plot, the line exhibits a jagged pattern with numerous spikes throughout the frequency range, with no single prominent peak. In contrast, the Welch plot shows a smoother line with a distinct spike around the 100 frequency marker.}''
Although the smoothing explanation is correct, 
\emph{both} plots have a notable spike around 100. This is particularly deceptive, suggesting the Welch method uniquely reveals a peak in the data without any other signs that the description is inaccurate.

In addition to the catchall ``deceptive'' error, we note that all other types except easily identifiable repetition, cutoff, grammar, or nonsense are also misleading to BVI users whose ground truth for charts may be the description.

\subsection{Error Prevalence in Generated descriptions}
\label{sec:manual_error_counts}
In this section, we compare the frequencies of error types across each of our generation strategies and datasets. 

For each dataset, we randomly select 100 figures (distinct from the 50 selected in \S\ref{sec:manual_error_types}) and manually label errors in their corresponding VL-T5 and MatplotAlt descriptions. We exclude the VL-T5 imageonly model, as we found that its captions never matched the given images. In total, we annotated 1000 figure descriptions. For each description we assign zero or more error types in order of priority based on the list in \S\ref{sec:manual_error_types}. If a caption states an incorrect value in a description of a trend but is otherwise correct, for example, we only assign it a value error, rather than value + trend or value + trend + deceptive errors. We label charts with no errors as ``\textbf{correct}'', and charts without factual errors (i.e. no value, identity, axis, trend, chart type, label, deceptive, or nonsense errors) as ``\textbf{value-correct}''. This means a blank caption would count as ``missing context'' under our criteria, but would still be value-correct. 

\begin{table*}[!htbp]
    \resizebox{\textwidth}{!}{%
    \begin{tabular}{c|l| M{0.06\linewidth}M{0.06\linewidth} |*{13}{M{0.045\linewidth}}}
\hline
\cellcolor[HTML]{FFFFFF}Dataset & Alt Text Type             & Correct                                                    & Value-Correct                                              & Chart Type                                               & Axis                                  & Value                                 &Ident-ity                                                & Trend                                 & Label                                                &  Missing Ctx.                                                & Unnec. Ctx.                                 & Repeti-tion                                                 & Cutoff                                                 & Number Name                        & Decept-ive                                                & Non-sense                                 \\ \hline \hline
VisText & VL-T5 - image + scene graph & 0 & \cellcolor[HTML]{87BDDC}43 & \textbf{0} & \cellcolor[HTML]{FFF2EB}\textbf{1} & \cellcolor[HTML]{FEE8DD}4 & \cellcolor[HTML]{F14432}{\color[HTML]{FFFFFF} 30} & \cellcolor[HTML]{FB7D5D}\textbf{24} & \cellcolor[HTML]{FFF2EB}\textbf{1} & \cellcolor[HTML]{000000}{\color[HTML]{FFFFFF} 95} & \cellcolor[HTML]{FFF2EB}\textbf{1} & \cellcolor[HTML]{FEE1D4}6 & \textbf{0} & \cellcolor[HTML]{FFF2EB}\textbf{1} & \textbf{0} & \cellcolor[HTML]{FCB89E}13 \\
VisText & VL-T5 - image + table & \cellcolor[HTML]{F1F7FD}\textbf{3} & \cellcolor[HTML]{4C99CA}{\color[HTML]{FFFFFF} \textbf{59}} & \textbf{0} & \cellcolor[HTML]{FFF2EB}\textbf{1} & \cellcolor[HTML]{FFF2EB}\textbf{1} & \cellcolor[HTML]{FCBEA6}\textbf{12} & \cellcolor[HTML]{FB7D5D}\textbf{24} & \cellcolor[HTML]{FFF2EB}\textbf{1} & \cellcolor[HTML]{000000}{\color[HTML]{FFFFFF} 95} & \cellcolor[HTML]{FFF2EB}\textbf{1} & \textbf{0} & \textbf{0} & \textbf{0} & \cellcolor[HTML]{FFEBE2}3 & \cellcolor[HTML]{FFEBE2}3 \\
\hline
VisText                         & Heuristic                 & \cellcolor[HTML]{73B2D7}\textbf{48}  & \cellcolor[HTML]{549ECD}{\color[HTML]{FFFFFF} \textbf{57}} & \cellcolor[HTML]{F14432}{\color[HTML]{FFFFFF} 30} & \textbf{0}                          & \textbf{0}                         & \cellcolor[HTML]{FEE1D4}6                         & \cellcolor[HTML]{FCA588}16         & \textbf{0}                                        & \cellcolor[HTML]{FCC4AD}11                        & \textbf{0}                         & \cellcolor[HTML]{FCC4AD}11                        & \textbf{0}                                        & \textbf{0}                & \textbf{0}                                        & \textbf{0}                         \\
VisText                         & turbo                     & \cellcolor[HTML]{E3EEF8}10          & \cellcolor[HTML]{D1E2F2}19           & \cellcolor[HTML]{FCAB8F}\textbf{15}               & \cellcolor[HTML]{FDCAB5}10          & \cellcolor[HTML]{FC8A6A}20         & \cellcolor[HTML]{EB372A}{\color[HTML]{FFFFFF} 32} & \cellcolor[HTML]{FB7D5D}22         & \cellcolor[HTML]{DE2B25}{\color[HTML]{FFFFFF} 34} & \cellcolor[HTML]{F14432}{\color[HTML]{FFFFFF} 30} & \cellcolor[HTML]{FFF2EB}\textbf{1} & \textbf{0}                                        & \cellcolor[HTML]{FEE8DD}4                         & \cellcolor[HTML]{FFEBE2}3 & \cellcolor[HTML]{FB7D5D}22                        & \textbf{0}                         \\
VisText                         & turbo + heuristic         & \cellcolor[HTML]{7FB8DA}45          & \cellcolor[HTML]{5DA4D0}{\color[HTML]{FFFFFF} 54}          & \cellcolor[HTML]{F5533B}{\color[HTML]{FFFFFF} 28} & \cellcolor[HTML]{FEE5D8}5           & \cellcolor[HTML]{FFF2EB}\textbf{1} & \cellcolor[HTML]{FEE8DD}\textbf{4}                & \cellcolor[HTML]{FCB296}14         & \cellcolor[HTML]{FFEBE2}3                         & \cellcolor[HTML]{FFF2EB}\textbf{1}                & \cellcolor[HTML]{FEE8DD}4          & \cellcolor[HTML]{FFEEE7}\textbf{2}                & \cellcolor[HTML]{FFF2EB}\textbf{1}                & \cellcolor[HTML]{FEDCCD}7 & \cellcolor[HTML]{FEE1D4}6                         & \textbf{0}                         \\
VisText                         & turbo + table             & \cellcolor[HTML]{ACD0E6}33          & \cellcolor[HTML]{61A7D2}{\color[HTML]{FFFFFF} 53}          & \cellcolor[HTML]{FC9D7F}\textbf{17}               & \cellcolor[HTML]{FCC4AD}11          & \cellcolor[HTML]{FEE8DD}4          & \cellcolor[HTML]{FDCAB5}10                         & \cellcolor[HTML]{FDD7C6}8          & \cellcolor[HTML]{FC8464}21                        & \cellcolor[HTML]{FDCAB5}10                        & \cellcolor[HTML]{FFF2EB}\textbf{1} & \cellcolor[HTML]{FFEEE7}\textbf{2}                & \cellcolor[HTML]{FDD7C6}8                         & \cellcolor[HTML]{FEE5D8}5 & \cellcolor[HTML]{FCAB8F}15                        & \cellcolor[HTML]{FEE5D8}5          \\
VisText                         & turbo + heuristic + table & \cellcolor[HTML]{6AADD5}{\color[HTML]{FFFFFF} \textbf{50}} & \cellcolor[HTML]{549ECD}{\color[HTML]{FFFFFF} \textbf{57}} & \cellcolor[HTML]{F96245}{\color[HTML]{FFFFFF} 26} & \cellcolor[HTML]{FEE5D8}5           & \textbf{0}                         & \cellcolor[HTML]{FFEEE7}\textbf{2}                & \cellcolor[HTML]{FCC4AD}11         & \textbf{0}                                        & \cellcolor[HTML]{FFF2EB}\textbf{1}                & \cellcolor[HTML]{FFEBE2}3          & \textbf{0}                                        & \cellcolor[HTML]{FEDCCD}7                         & \cellcolor[HTML]{FEE5D8}5 & \cellcolor[HTML]{FEE5D8}5                         & \cellcolor[HTML]{FFEBE2}3          \\ \hline
Gallery              & Heuristic                 & \cellcolor[HTML]{E3EEF8}10          & \cellcolor[HTML]{3B8BC2}{\color[HTML]{FFFFFF} \textbf{65}} & \cellcolor[HTML]{FEE1D4}6                         & \cellcolor[HTML]{FDD0BC}\textbf{9}  & \cellcolor[HTML]{FDD7C6}\textbf{8} & \cellcolor[HTML]{FC8A6A}20                        & \cellcolor[HTML]{FEE8DD}\textbf{4} & \textbf{0}                                         & \cellcolor[HTML]{67000D}{\color[HTML]{FFFFFF} 67} & \textbf{0}                         & \cellcolor[HTML]{FB694A}{\color[HTML]{FFFFFF} 25} & \cellcolor[HTML]{FFF2EB}\textbf{1}                & \textbf{0}                & \cellcolor[HTML]{FFF2EB}\textbf{1}                & \cellcolor[HTML]{FFF2EB}\textbf{1} \\
Gallery              & turbo                     & \cellcolor[HTML]{ACD0E6}33          & \cellcolor[HTML]{7BB7D9}46            & \cellcolor[HTML]{FDD0BC}9                         & \cellcolor[HTML]{FDCAB5}\textbf{10} & \cellcolor[HTML]{FEDCCD}\textbf{7} & \cellcolor[HTML]{FDD0BC}\textbf{9}                & \cellcolor[HTML]{FFEBE2}\textbf{3} & \cellcolor[HTML]{FEE8DD}4                         & \cellcolor[HTML]{FEE1D4}6                         & \cellcolor[HTML]{FEE5D8}5          & \textbf{0}                                        & \cellcolor[HTML]{FC8A6A}20                        & \textbf{0}                & \cellcolor[HTML]{F03D2D}{\color[HTML]{FFFFFF} 31} & \textbf{0}                         \\
Gallery              & turbo + heuristic         & \cellcolor[HTML]{98C7DF}\textbf{39} & \cellcolor[HTML]{4C99CA}{\color[HTML]{FFFFFF} 59}          & \cellcolor[HTML]{FFF2EB}\textbf{1}                & \cellcolor[HTML]{FDD0BC}\textbf{9}  & \cellcolor[HTML]{FEE1D4}\textbf{6} & \cellcolor[HTML]{FC8A6A}20                        & \cellcolor[HTML]{FEE8DD}\textbf{4} & \cellcolor[HTML]{FFF2EB}\textbf{1}                & \cellcolor[HTML]{FFEBE2}\textbf{3}                & \cellcolor[HTML]{FEE8DD}4          & \textbf{0}                                        & \cellcolor[HTML]{DE2B25}{\color[HTML]{FFFFFF} 34} & \textbf{0}                & \cellcolor[HTML]{FCB296}14                        & \cellcolor[HTML]{FFEEE7}\textbf{2} \\ \hline
\end{tabular}}
    \caption{\textbf{MatplotAlt's heuristic and VLM methods can produce factually correct and informative long-form alt text, but VLMs still struggle without extra supervision about chart data and features.} For 100 descriptions from each generation strategy and dataset, this table shows the number labeled as correct and value-correct (darker blue is better), and the number containing each type of error (darker red is worse).}
    \label{table:error_counts}
\end{table*}

Table \ref{table:error_counts} records the number and type of errors in the 100 captions from each dataset and generation strategy. We find that turbo + heuristic and turbo + table + heuristic produce more accurate alt texts than GPT4-turbo alone, with lower counts of value ($20 \rightarrow 1$), identity ($32 \rightarrow 4$), trend ($22 \rightarrow 14$), label ($34 \rightarrow 3$), missing context ($30 \rightarrow 1$), and deceptive errors ($22 \rightarrow 6$). At the same time, errors in heuristic alt text tend to carry over to turbo + heuristic and turbo + heuristic + table descriptions, even when they could be corrected with the given datatable or image. For example, almost all of the chart type errors in turbo + heuristic / turbo + table + heuristic originated with MatplotAlt's lack of support for area charts. Excluding these chart type errors specifically on area plots, the number of correct turbo + heuristic descriptions rises to 61 and turbo + heuristic + table increases to 65. Similarly, at the time of annotation, most trend errors in turbo + heuristic captions were carried over from our simple heuristics for detecting trends.

Adding markdown data tables to prompts also increased the accuracy of descriptions, with smaller improvements in most error types, and the lowest number of trend errors out of all GPT4 generation methods. Qualitatively, we observed turbo + table and turbo + table + heuristic generations to have more detailed trend descriptions, often mentioning notable peaks and troughs outside the global min / max and periods of fluctuation not captured by other methods. 

Overall, of the turbo methods, turbo + heuristic + table had the most correct captions (\textbf{50}) followed by turbo + heuristic (45), turbo + table (33), and turbo (10). This trend indicates that adding extra supervision to GPT4-turbo prompts reduces the number of deceptive factual errors in generated descriptions. Turbo's low performance is also notable, with only 10 of its captions containing no errors, and 81/100 containing some kind of factual inaccuracy. This suggests that, out of the box, GPT4 and possibly other large VLMs have improved at generating \emph{fluent}, but not necessarily \emph{correct}, alt text for visualizations.

With the smaller finetuned VL-T5 models, captions tended to be more broad in addition to shorter, usually only including a line or two about general increasing and decreasing trends. We found that only 5 captions from each VL-T5 + data strategy contained mentions of specific data values, and only 3 of these from image + datatable were factually correct. This disqualified all other captions under our criteria due to missing context. 43 and 59 of the scenegraph and datatable VL-T5 captions were value-correct respectively, mostly due to omission. These results highlight the difficulty of our qualitative evaluation. In addition to not containing hallucinations, we expect heuristic and turbo alt texts to explicitly reference underlying chart data. We note that many crowdsourced L3 captions in VisText would also be labeled as missing context.

There were a lower overall number of correct captions for Matplotlib gallery figures. Our heuristic alt text in particular tended to break down in this setting due to its lack of ability to handle complex inputs, with over two thirds of its captions lacking necessary context. Mirroring our VisText results, however, heuristics still helped reduce deceptive errors in turbo + heuristic, which had the highest number of correct descriptions (39/100). This indicates that it is helpful to add heuristics in both simple univariate settings and complex ones where parsing the chart into a data table is infeasible. 

Interestingly, we also labeled more of turbo's descriptions as correct on gallery figures compared to VisText, largely due to fewer identity and trend errors. These decreases are likely because there are similarly fewer clear trends and statistics to describe in gallery figures. The content of VisText charts may have also overlapped with GPT4-turbo's pretraining data in ways that caused identity errors. In several VisText captions, for example, turbo incorrectly labels 2008/2009 as a minimum and cites the economic crisis. This suggests that large VLMs may counter-intuitively be \emph{less} likely to hallucinate specific data points on out-of-distribution figures, although more study is needed to confirm this claim.

%% file: content/discussion.tex
\section{Discussion}
\label{sec:discussion}
We present MatplotAlt, a Python library to facilitate adding alt text to Matplotlib figures in computational notebooks. Our design decisions to generate and surface alt text for Matplotlib figures are informed by prior work documenting barriers BVI users face in accessing visualization tools \cite{potluri2023notably, srinivasan2023azimuth, sharif2021understanding}, and by previous systems for automatically generating alt text \cite{masry2023unichart, sharif2022voxlens, lundgard2021accessible, sharif2018evographs}. Our quantitative evaluation of the descriptions shows that MatplotAlt's heuristic-based descriptions can adhere to the style of human-written L3 alt texts, and that prompting VLMs with heuristic descriptions can be used to steer their generations closer to ground-truth human captions without significantly harming their ability to describe complex figures. Similarly, we observe that prompting with heuristic alt text is a simple way to reduce hallucinations and improve the factuality of VLM-generated figure descriptions, even when chart data cannot be easily formatted in text.
We provide recommendations to generate and surface effective image descriptions, and discuss the implications of automatically generating alt-text for BVI and sighted notebook authors. We close with limitations of our study.
\subsection{Recommendations to generate alt text}
Based on our evaluation and design considerations, we provide the following recommendations for generating alt text with MatplotAlt: 
\begin{itemize}
    \item We suggest users replace \texttt{plt.show} with \texttt{show\_with\_alt} or \texttt{show\_with\_api\_alt} depending on context. Using heuristic alt text in domain-specific notebooks aimed at knowledgeable audiences may be desirable as they prefer concise, accurate descriptions. In more general notebooks for personal data analysis, extra VLM context may be preferred~\cite{Seo:2024:MAIDRAI}.
    \item Our evaluation shows that prompting models with heuristic alt text and data tables increases the factuality of generations, can steer captions closer to human-written descriptions, and does not significantly harm similarity between descriptions and figures. As a result, we include heuristics and tables when possible by default in \texttt{show\_with\_api\_alt}.
    \item Using the HTML option to surface alt text may lead to a better user experience for notebooks that are published on the web. As the accessibility of notebook interfaces is evolving, using the options to embed descriptions as code comments or markdown cell content may make these notebooks accessible to BVI users who experience barriers with notebook interfaces. Saving descriptions into image or text files, on the other hand, may make it easy to use them alongside corresponding figures in other contexts such as research publications.
\end{itemize}

\subsection{Implications of Automating Alt Text Generation}
Matplotalt provides a variety of options to generate and surface alternative texts for figures with no human input. Using these tools, BVI experts can perceive notebook visualizations that they may encounter in the wild to learn new concepts or conduct data exploration. At the same time, MatplotAlt's current description generation methods are error-prone and could be misleading when used by BVI notebook users, as seen in our evaluation of description correctness (\S\ref{sec:manual_error_counts}). We observed that the BVI research collaborator on the team was more punishing of error occurrences as he felt they mislead him on the insights he would expect to derive from the data. Future efforts could study the applicability of VLM-generated descriptions and the trust that BVI experts place in these potentially error-prone, very detailed descriptions. Additional validation support, such as access to the original data or the ability to have a sighted collaborator check results, would be helpful here. In other words, \emph{what is the detail-accuracy tradeoff that BVI experts prefer in descriptions of data visualizations?}

To alleviate accessibility barriers posed by notebook environments, MatplotAlt provides command line tools to generate and export notebooks with described Matplotlib images. This functionality has the potential for the system to be integrated into testing and quality control processes that developers follow. For example, MatplotAlt could be added to existing git pipelines that authors use to test notebooks.  
\subsection{Implications for interactive data exploration}
Several efforts  equip BVI users with the tools to interactively explore data~\cite{Potluri:2022:PSST,Thompson:2023:ChartReader}. MatplotAlt is complimentary to these; it provides a standardized mode to access alternative text and requires little to no change in notebook hosting environments. Future efforts could add question-answering abilities to MatplotAlt, making the data exploration process for BVI users iterable. Similarly, MatplotAlt's generation mechanisms could be integrated into existing data exploration tools such as MAIDR and Umwelt~\cite{Seo:2024:MAIDR,Seo:2024:MAIDRAI,Zong:2024:Umwelt}. The approaches explored in MatplotAlt to use underlying data to enhance descriptions could inform future efforts to create interactive sonifications to help BVI users answer their questions.

\subsection{Limitations}
\label{sec:limitations}
MatplotAlt supports multiple ways of generating and surfacing alt text for Matplotlib figures. Alternative text and figure descriptions alone, however, do not support the full extent of data exploration needs of BVI experts\cite{sharif2021understanding, sharif2022voxlens}. We believe that alternative text and text descriptions, however, serve as a standardized way of exposing accessible graphics across a variety of scenarios and serve as an important first step. Future versions of MatplotAlt could offer functionality for developers to interact with the data via Q and A  to support interactive explorations of data by integrating with systems like MAIDR~\cite{Seo:2024:MAIDRAI}.

Our evaluations of alt text generations may not be perceived as rigorous as they do not offer definitive guidelines of specific generation strategies of MatplotAlt as good defaults. Given the variability in alternative text quality across different measures as shown in our evaluations, and varying preferences of BVI users on alt text~\cite{Seo:2024:MAIDRAI}, providing definitive defaults that work for all contexts is not feasible. We trust the developers producing notebooks with Matplotlib figures to use their best judgment and expertise to decide on a good alt-text generation strategy. Although the \texttt{desc\_level} parameter allows users to modify the inclusion of information, we leave it to feature work to support changing the verbosity and ordering of generated descriptions, which are also important modes of customization for BVI users \cite{jones2024customization}.
 
Additionally, our generation strategies may not be effective for several chart types or plotting methods used by Matplotlib users as our evaluations used carefully curated images. Future work could programmatically evaluate our methods on a larger subset of the Jetbrains dataset~\cite{Guzharina_undated-ai}, which may be more representative of MatplotAlt use in-the-wild. The \texttt{alttextify} command begins to make the infrastructure to run such at-scale evaluations available, but still requires executing arbitrary code, which is potentially dangerous in untrusted notebooks.


\subsection{Conclusion}
We present MatplotAlt, a Python library to add alt text to Matplotlib figures in computational notebooks. MatplotAlt implements functions to generate and embed or export figure descriptions in a single line of code or command. Our initial qualitative evaluations indicate that MatplotAlt can generate effective alt text for Matplotlib figures, and that heuristic alt text and data tables can be used to increase the factuality of VLM descriptions. Our goal is that this work will eventually see adoption and start increasing the percentage of notebooks on the web that are accessible to BVI users. We also hope that our project will gain collaborators (including from the Matplotlib and Jupyter teams) who contribute their experience to improve its accessibility and generations.






%% file: content/acknowledgements.tex
\section{Acknowledgements}
This work was supported by GitHub, Microsoft, and Google. The contents of this article were developed under a grant from the National Institute on Disability, Independent Living, and Rehabilitation Research (NIDILRR grant number 90REGE0026-01-00) funding the Center for Research and Education on Accessible Technology and Experiences (CREATE). NIDILRR is a Center within the Administration for Community Living (ACL), Department of Health and Human Services (HHS). The contents of this article do not necessarily represent the policy of NIDILRR, ACL, HHS, and you should not assume endorsement by the Federal Government. Open Access funding enabled by the University of Michigan School of Information.
We thank members of the Make4All lab and visualization system seminar at the University of Washington for their valuable feedback.

%% file: supp_content/example_gallery.tex
\section{Example Gallery}
\label{sec:example_gallery}
\begin{center}
    \begin{figure*}
    \resizebox{\textwidth}{!}{%
        \begin{subfigure}[t]{2.8in}
            \includegraphics[alt={A horizontal bar chart with month on the y-axis and average number of bikes crossing the fremont bridge on the x-axis ranging from 0 to 250000. The number of bikes increases steadily from a min around 100000 in December to a peak around 240000 in July, decreasing back down to near the min in January.},scale=0.4]{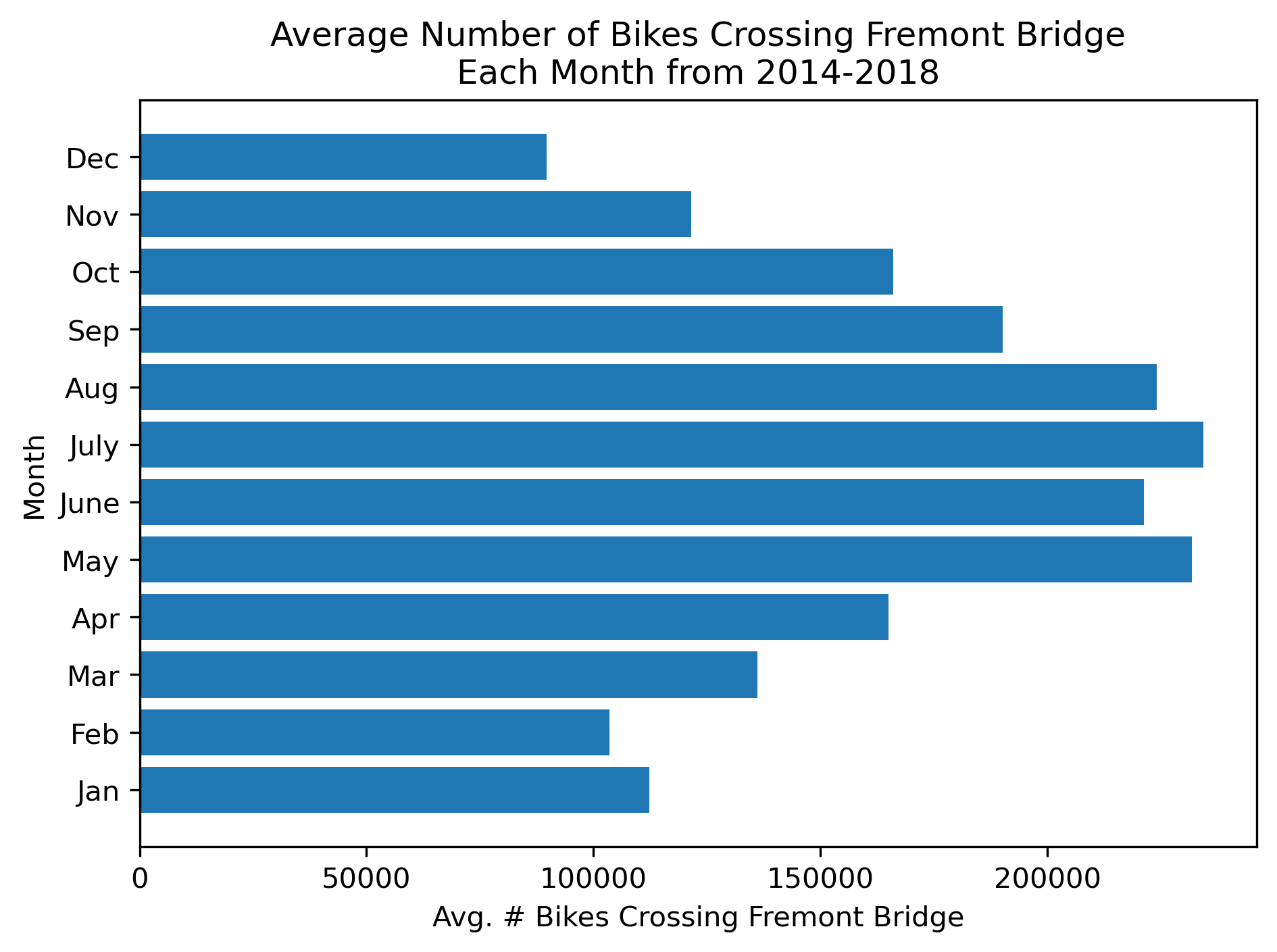}
        \end{subfigure}\hfill
        \begin{subfigure}[t]{2.8in}
            \includegraphics[alt={A scatterplot depicting two 2d gaussian distributions labeled "the blue dots" and "the orange dots" which are plotted in their respective colors. The x-axis ranges from -10 to 8 and the y-axis ranges from -3 to 8. The orange dots appear to be centered around (-5, 3) and the blue dots are centered around approximately (2, 2).},scale=0.1]{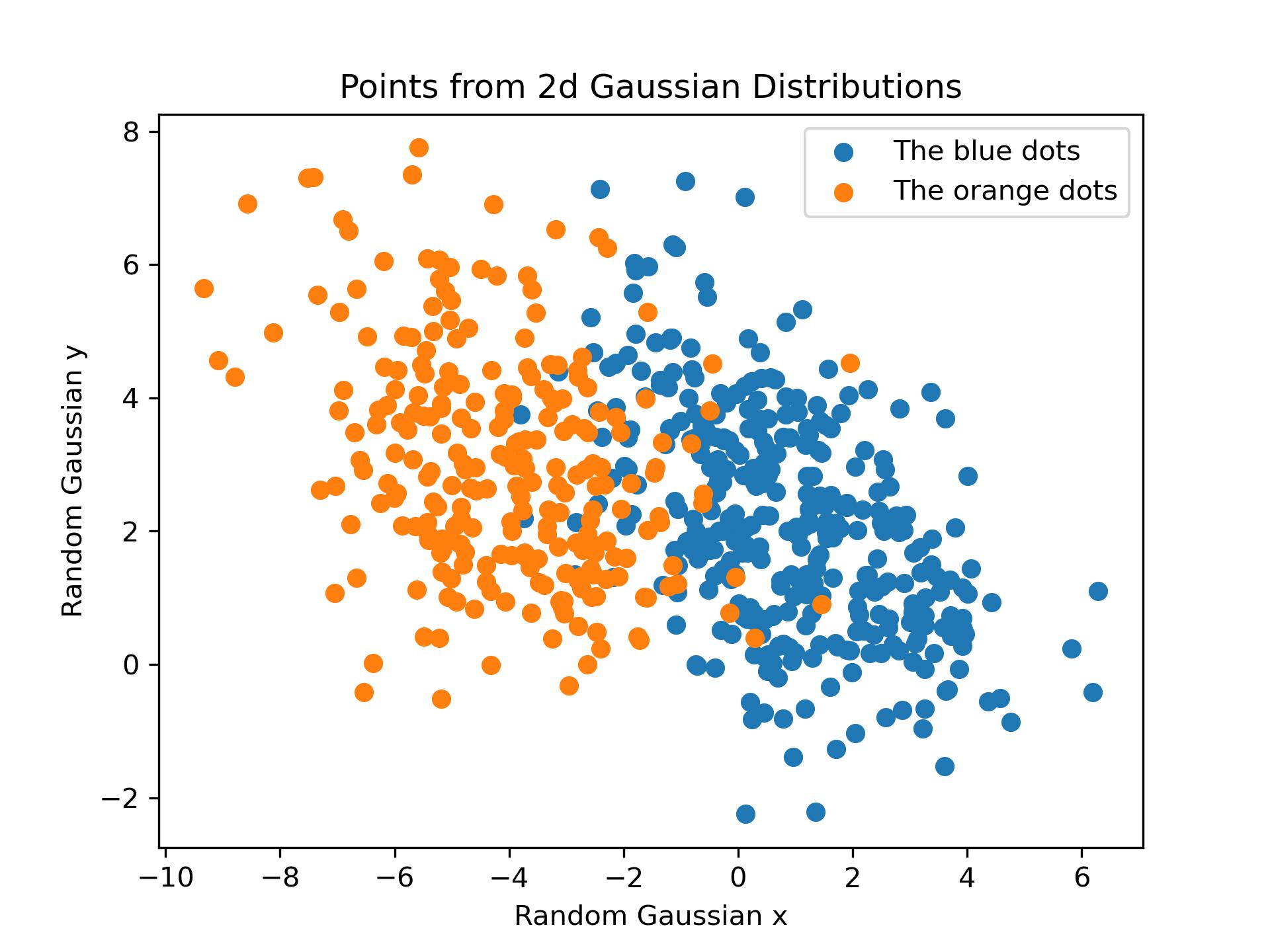}
        \end{subfigure}\hfill
        \begin{subfigure}[t]{2.8in}
            \includegraphics[alt={Four line plots with the suptitle "Anscombe's quartet". Each line plot has the same mean, standard deviation, and line of fit, but each has a distinct pattern. The top left chart is a fluctuating upward trend starting from around (4, 4) and increasing to around (15, 10). The top right chart is a parabola starting from (4, 4) and peaking around (10, 11) before decreasing. The points in the bottom left chart all lie exactly on a line from around (4, 5) to (15, 8), with a single outlier at around (12.5, 13). Finally, the bottom right chart's points all share an x-value of 7.5 with varying y values, except a single outlier at around (20, 12).},scale=0.1]{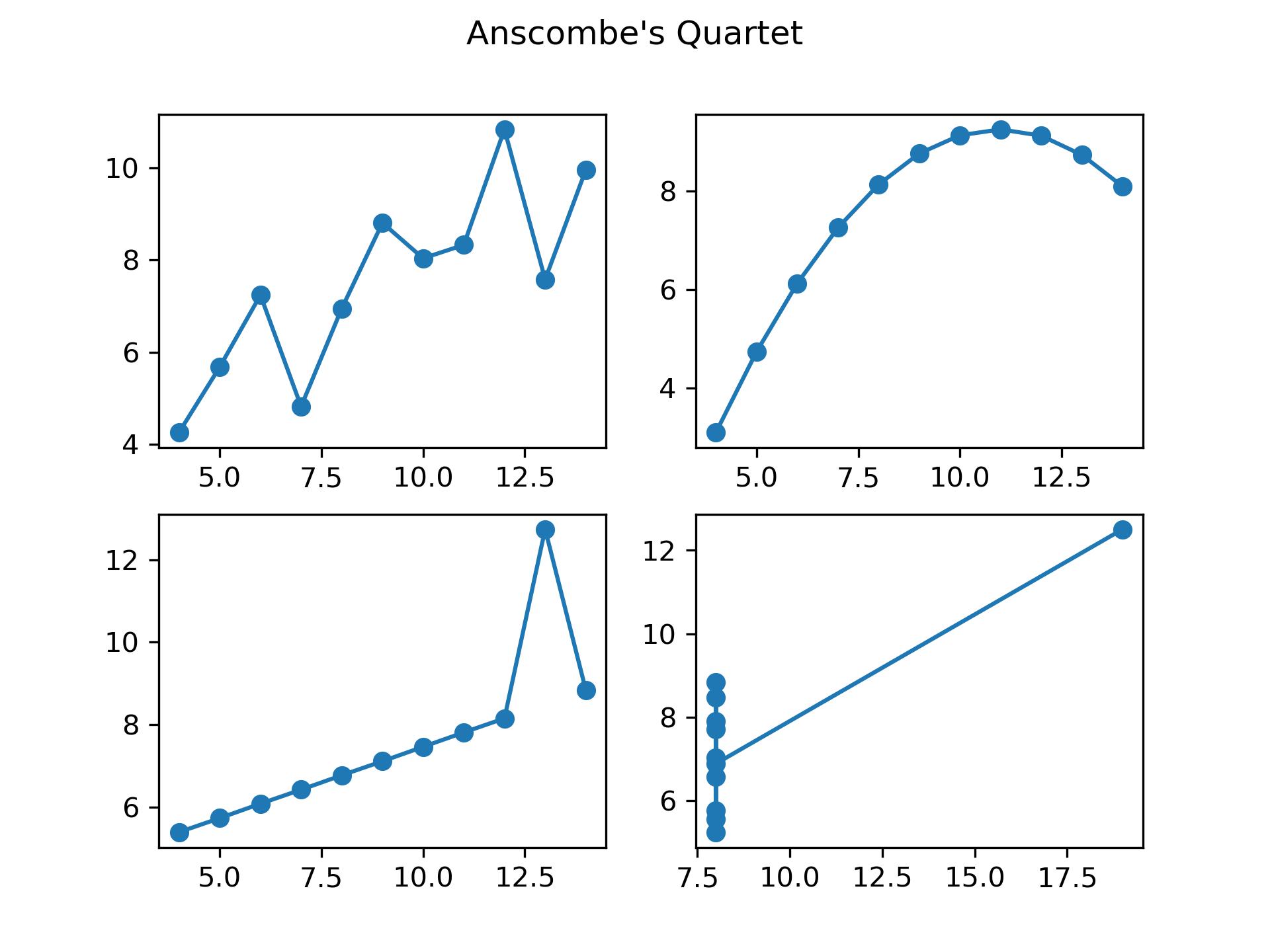}
        \end{subfigure}} \\
    \resizebox{\textwidth}{!}{%
        \begin{subfigure}[t]{2.8in}
            \includegraphics[alt={A line chart plotting the number of sunshine hours in Seattle (in orange) vs. the number of bikes that cross fremont bridge each month (in blue). Both lines appear to be correlated, starting near a low in January, peaking in July, and then decreasing back to near the min in december. An arrow points to the peak of the bikes line indicating there were 234421 bikes in July.},scale=0.4]{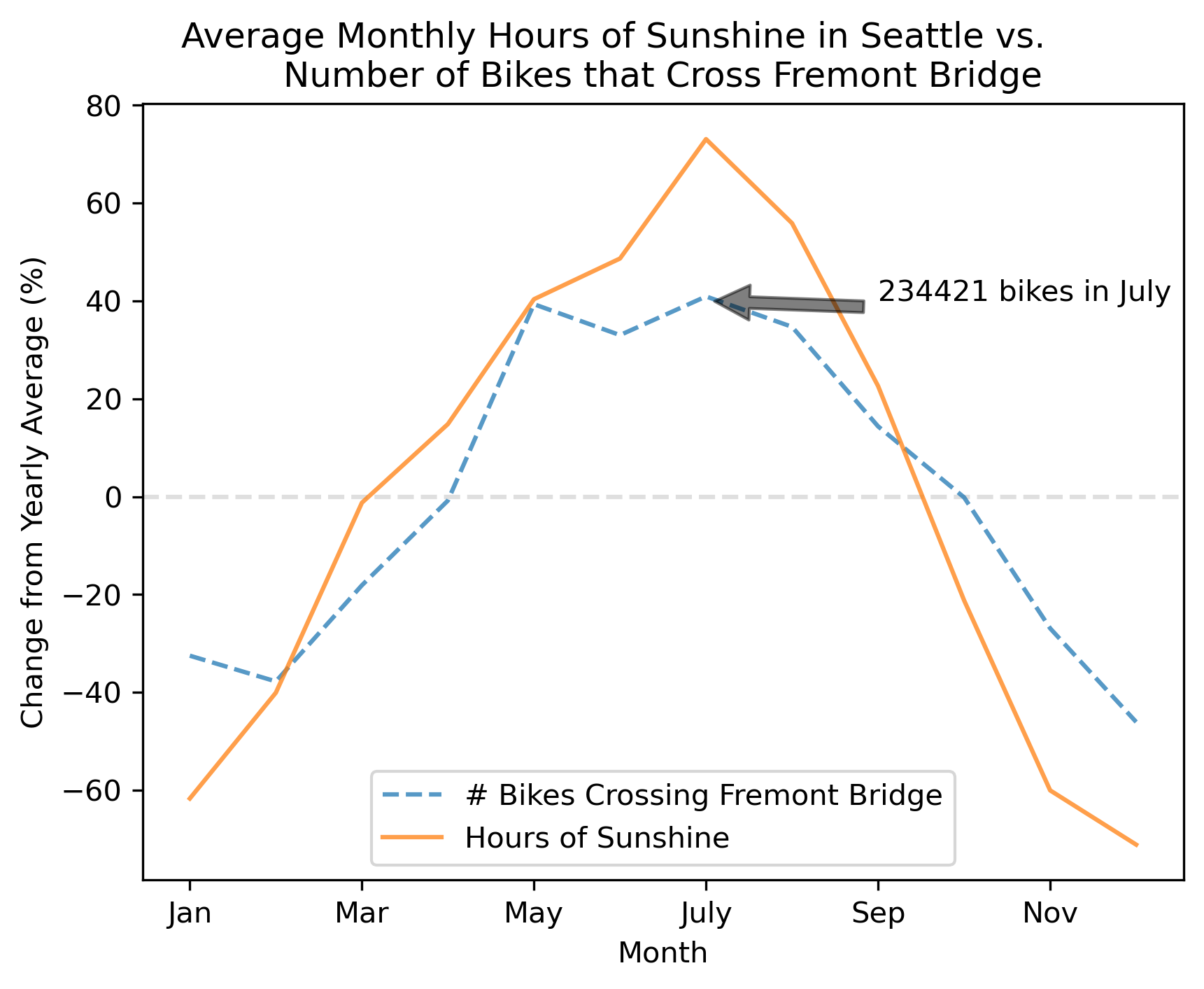}
        \end{subfigure}\hfill
        \begin{subfigure}[t]{2.8in}
            \includegraphics[alt={A horizontal strip plot depicting the distribution of x-values for the 2d gaussian distributions in the scatterplot above. The blue dots are on top, and the orange dots are plotted below. The strips reveal that the blue dots are centered around approximately x=1, and the orange dots are centered around about x=-4},scale=0.23]{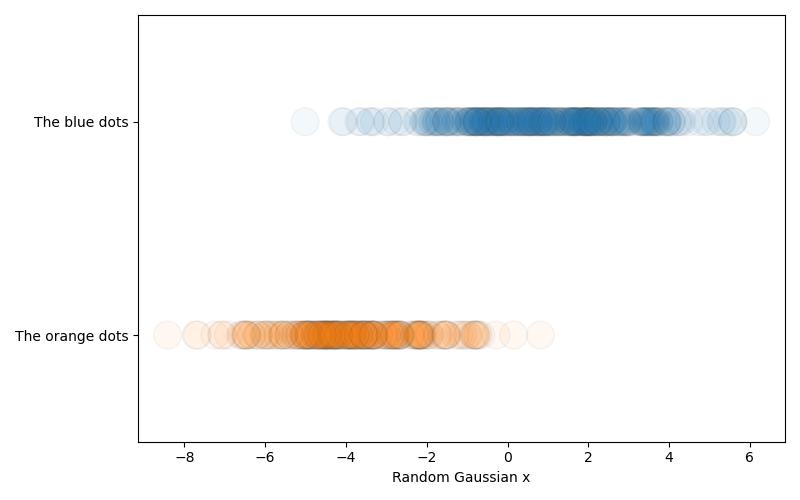}
        \end{subfigure}\hfill
        \begin{subfigure}[t]{2.8in}
            \includegraphics[alt={A vertical boxplot with four boxes depicting the distribution of y-values in each of the charts from Anscombe's quartet. Despite having the same line of fit, each box has a distinct distribution. The box for the first plot (colored blue) has a median around y=7 with a large interquartile range from around y=4 to y=11. The second box (in orange) has a median around 8 with an interquartile range from around 5 to 9. The third box (in green) has a median around 7, an interquartile range from 5 to 9, and a single outlier around y=13. Finally, the fourth boxplot (in red) has a similar interquartile range and median to the third, with an outlier around y=12.},scale=0.1]{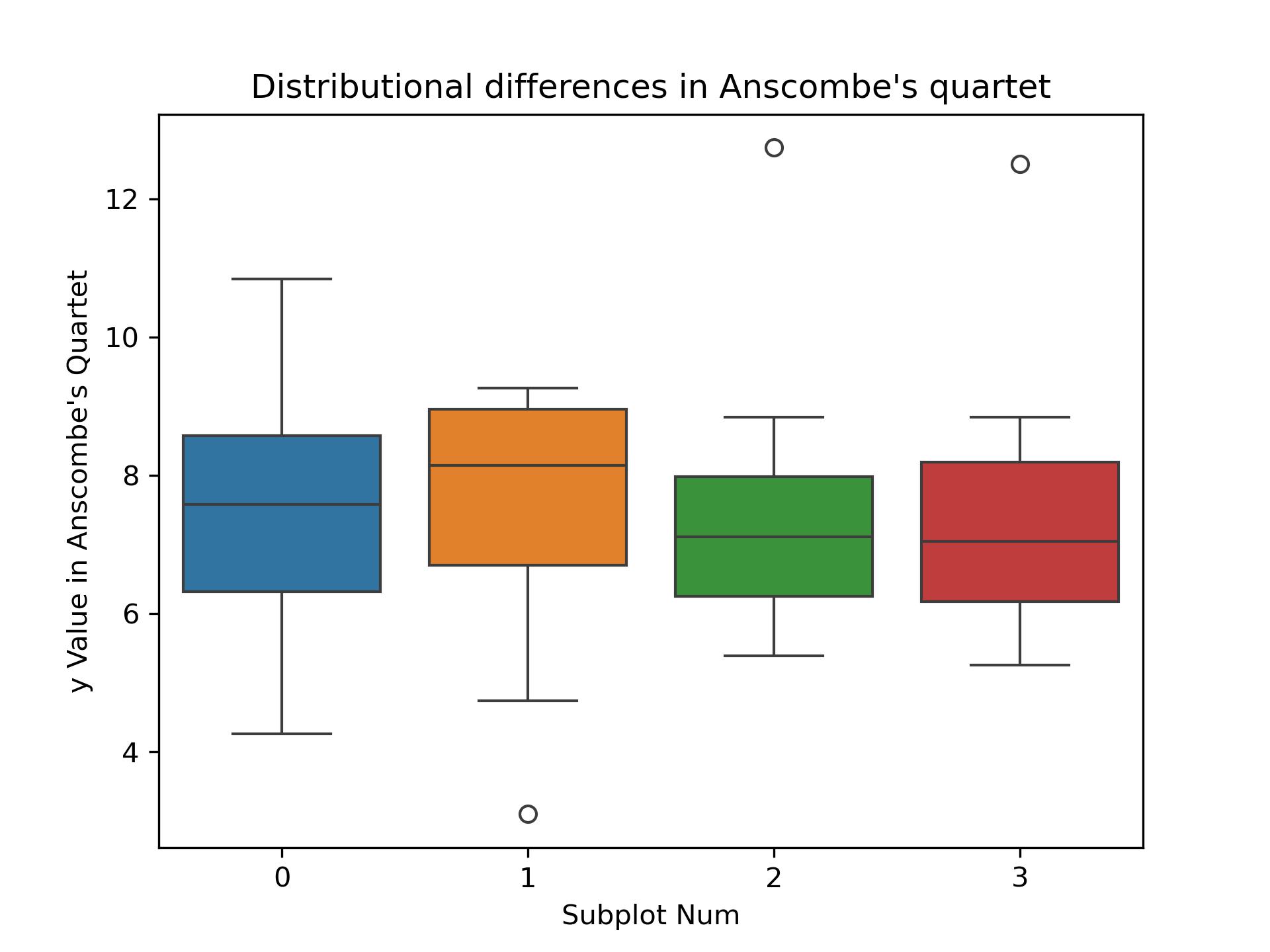}
        \end{subfigure}}\\
    \resizebox{\textwidth}{!}{%
        \begin{subfigure}[t]{2.8in}
            \includegraphics[alt={A radial line chart depicting the number of sunshine hours each month in seattle. The months are depicted every 30 degrees ranging from July on the left to January on the right. The number of sunshine hours are depicted on each radial line starting from 0 in the center to a max of 312 on the outermost line. The hours of sunshine are plotted in a filled orange line, peaking around 312 in July, then decreasing to a min around 50 in December before increasing again.},scale=0.09]{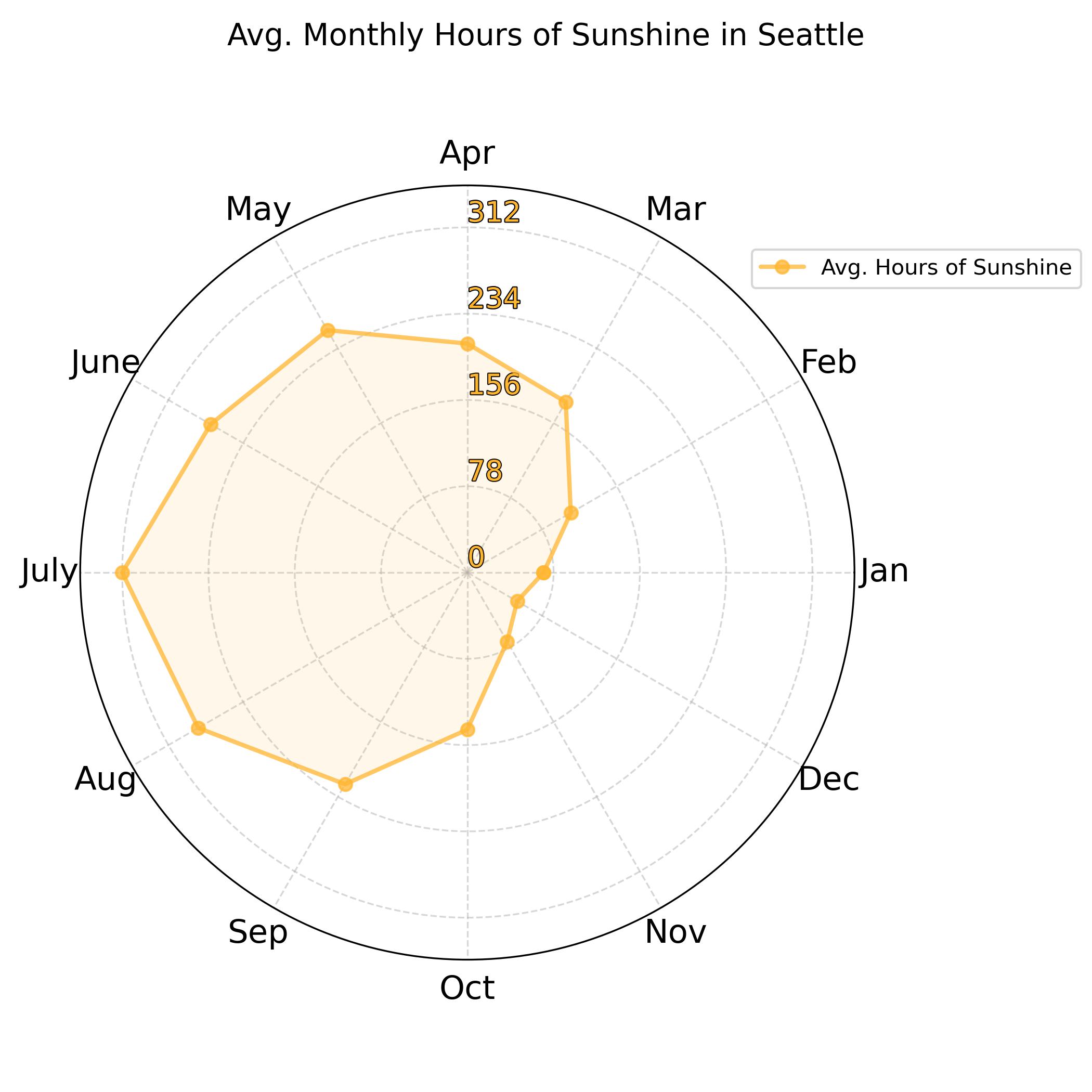}
        \end{subfigure}
        \begin{subfigure}[t]{2.8in}
            \includegraphics[alt={A 7x6 heatmap depicting the binned number of points from the same 2d gaussian distributions in each 2x2 square. The x-axis ranges from -8 to 6 and the y-axis ranges from -4 to 8. There is a colorbar showing that the minimum number of points (0) is plotted in dark blue, and the maximum number of points (60) is plotted in bright yellow. The square containing the max number of points is at approximately (1,1), and there are several bins containing 0 points around the edges of the plot. The points appear to be centered near the center of the heatmap, with notable peaks around (1, 1) and (-3, 3).},scale=0.34]{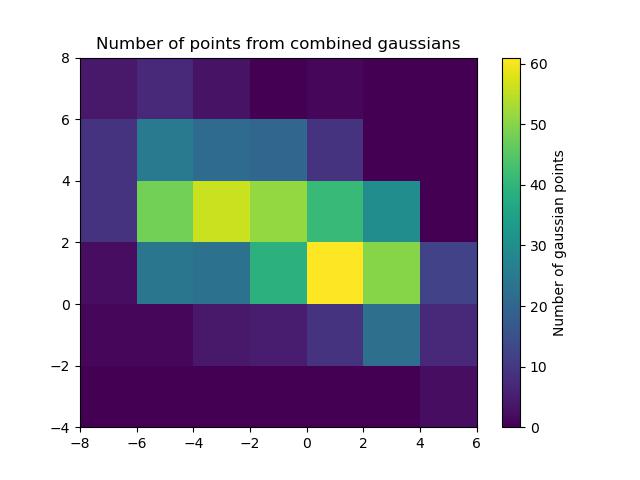}
        \end{subfigure}
        \begin{subfigure}[t]{2.8in}
            \includegraphics[alt={An image depicting the fractal z^3 + 0.66 -- where each pixel corresponds to the number of iterations each complex point at the position x + yi takes to reach a threshold of 2. A colorbar indicates the number of iterations ranging from 0 in dark blue to 17 in bright yellow. The plot appears to have three-fold symmetry centered around the center, with the max number of iterations occurring in intricate sections around the top middle, bottom left, and bottom right regions.},scale=0.42]{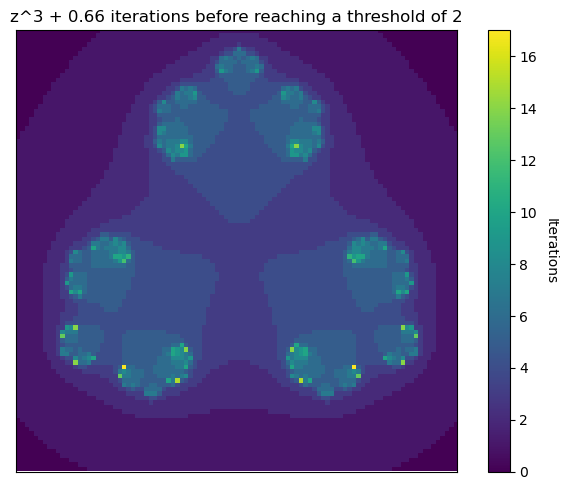}
        \end{subfigure}}\\
    \resizebox{\textwidth}{!}{%
        \begin{subfigure}[t]{2.8in}
            \includegraphics[alt={A pie chart depicting the same sunshine hours datset. Each slice has a distinct color, a text label in the slice indicating the month, and a number on the outside indicating the percentage of annual sunshine each month contributes. The max occurs in July with 14.4\% of annual sunshine, and the min occurs in December at only 2.4\%. The data decrease from July to december, before increasing again.},scale=0.5]{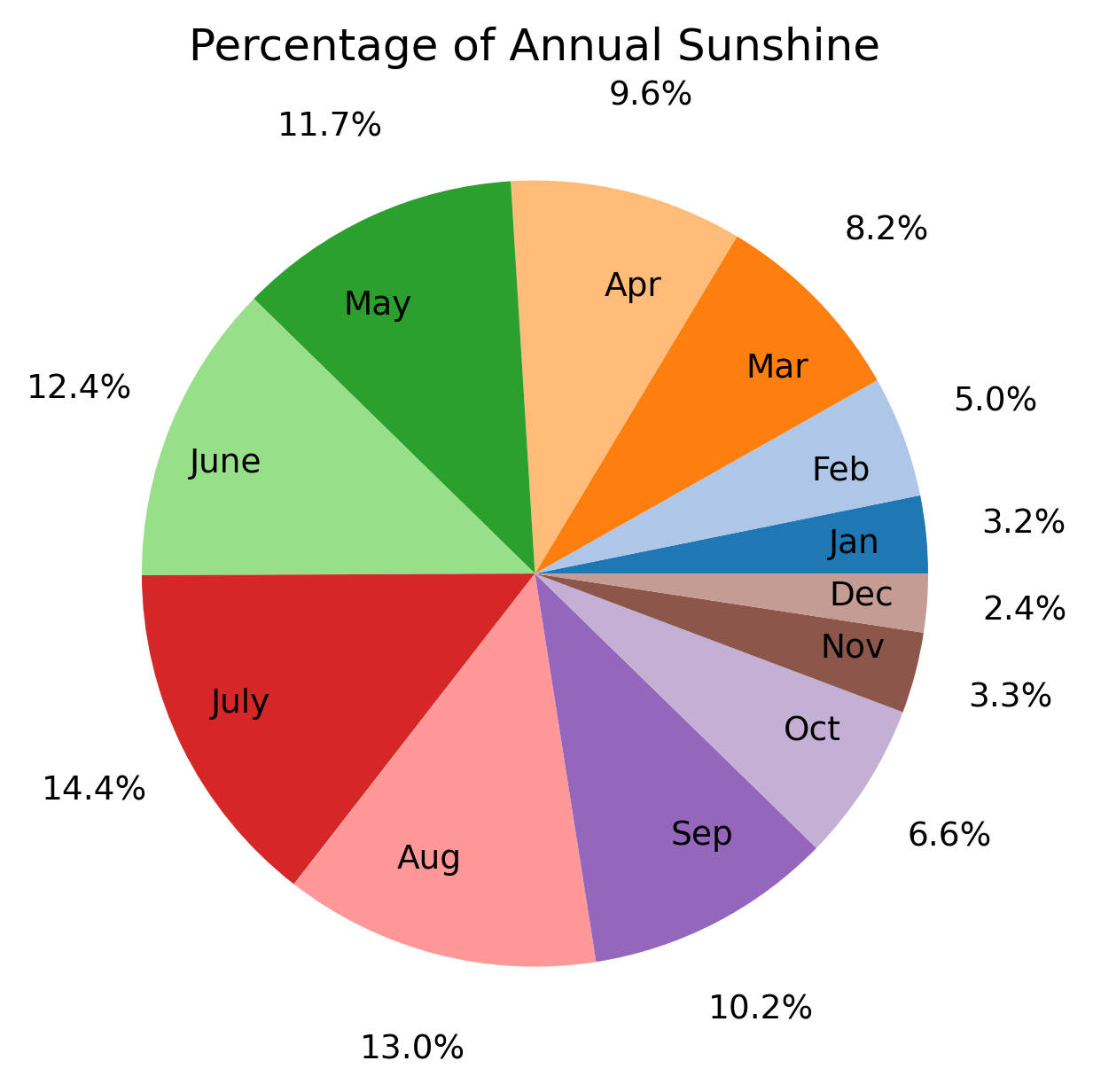}
        \end{subfigure}
        \begin{subfigure}[t]{2.8in}
            \includegraphics[alt={A contour plot depicting the a similar 2d gaussian distribution as the heatmap above titled "Number of points from combined gaussians". There appear to be 7 contour lines with values 0, 8, 16, 24, 32, 40, and 48 respectively. The max contour is centered around approximately (-5, 5). there is another contour near the max at approximately (-3, 0). The contours decrease radiating out from these two centers.},scale=0.3]{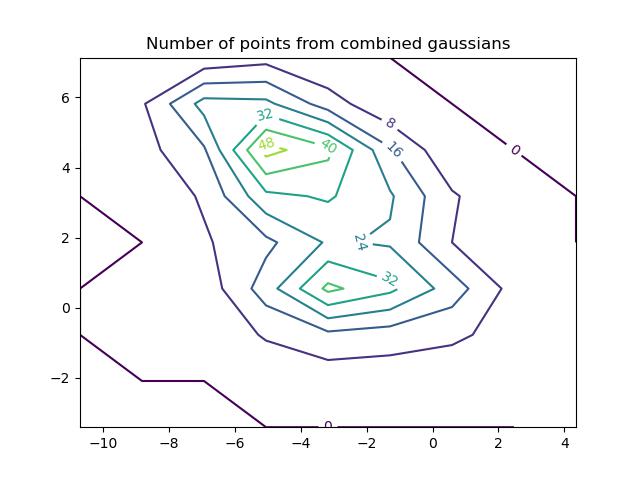}
        \end{subfigure}
        \begin{subfigure}[t]{2.8in}
            \includegraphics[alt={A blank plot titled "blank".},scale=0.42]{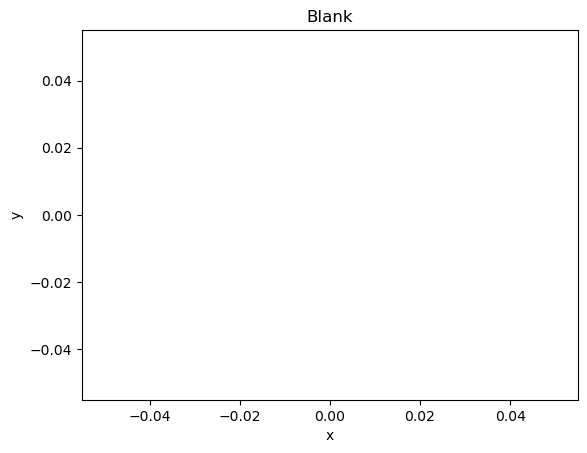}
        \end{subfigure}}
        \caption{Types of charts supported by MatplotAlt. From top left to bottom right: a bar plot, a scatter plot, multiple subplots, a line plot, a strip plot, a boxplot, a radial line plot, a heatmap, an image displayed with \texttt{imshow}, a pie chart, a contour plot, and a blank plot. Example heuristic alt texts for each of the charts are listed in Table \ref{tab:example_captions}.}
        \label{tab:example_gallery}
    \end{figure*}
\end{center}

We provide a gallery of MatplotAlt's supported chart types in Table \ref{tab:example_gallery}, a list of their corresponding heuristic-based alt texts in Table \ref{tab:example_captions}, and an example data table created using \texttt{generate\_alt\_text} in Table \ref{tab:example_md_datatable}. We also compare each of our alt text generation methods on two charts in Tables \ref{table:example_descriptions1} and \ref{table:example_descriptions2}, with sections labeled by L1-L3 semantic level. Finally, we provide two examples from VisText and the Matplotlib gallery, labeled with a range of different error types in Table \ref{table:example_error_descriptions}.

\begin{center}
    \begin{table*}
        \begin{tabular}{l|p{0.8\textwidth}}
            \toprule
            \textbf{Plot type} & \textbf{Example Heuristic-based Alt Text} \\
            \midrule
            \textbf{Bar plot} & A bar chart titled 'average number of bikes crossing fremont bridge each month from 2014-2018'. Avg. \# bikes crossing fremont bridge is plotted on the x-axis from 0 to 250000 using a linear scale and month is plotted on the y-axis from jan to dec using a datetime scale. The data has 12 points, a minimum value of x=89700 at y=dec, a maximum value of x=234400 at y=july, and an average of x=166300. \\
            \midrule
            \textbf{Line plot} & A line plot titled 'average monthly hours of sunshine in seattle vs. Number of bikes that cross fremont bridge'. Month is plotted on the x-axis from jan to nov using a datetime scale and change from yearly average (\%) is plotted on the y-axis from -80 to 100 using a linear scale. \# bikes crossing fremont bridge is plotted in dark blue and hours of sunshine is plotted in orange. There is a horizontal line at y=0.0. An annotation reads '234421 bikes in july'. \# bikes crossing fremont bridge has a minimum value of y=-46.08 at x=12, a maximum value of y=40.93 at x=7, and an average of y=0. Hours of sunshine has a minimum value of y=-71.15 at x=12, a maximum value of y=73.09 at x=7, and an average of y=0.\\
            \midrule
            \textbf{Pie chart} & A pie chart titled 'percentage of annual sunshine'. There are 12 slices: jan (3.19\%), feb (4.993\%), mar (8.229\%), apr (9.57\%), may (11.7\%), june (12.39\%), july (14.42\%), aug (12.99\%), sep (10.22\%), oct (6.565\%), nov (3.329\%), and dec (2.404\%). The data has a standard deviation of x=4.006, an average of x=8.333, a maximum value of x=14.42, and a minimum value of x=2.404. The data strictly increase up to their max at x=14.42, then strictly decrease.\\
            \midrule
            \textbf{Radial line chart} & A radial line plot titled 'avg. Monthly hours of sunshine in seattle'. The x-axis ranges from jan to dec using a datetime scale and the y-axis ranges from 0 to 312 using a linear scale. Avg. Hours of sunshine is plotted in orange. Avg. Hours of sunshine has a minimum value of y=52 at x=dec, a maximum value of y=312 at x=july, and an average of y=180.2.\\
            \midrule
            \textbf{Scatter plot} & A scatter plot titled 'points from 2d gaussian distributions'. Random gaussian x is plotted on the x-axis from -12.5 to 7.5 and random gaussian y is plotted on the y-axis from -4 to 10, both using linear scales. The blue dots is plotted in dark blue and the orange dots is plotted in orange. The blue dots has 350 points, an average of x=1.149, an average of y=1.915, a linear fit of y=-0.5268x+2.52, and 6 outliers. The orange dots has 250 points, an average of x=-4.114, an average of y=3.055, a linear fit of y=-0.3893x+1.453, and 5 outliers.\\
            \midrule
            \textbf{Strip plot} & A strip plot. Random gaussian x is plotted on the x-axis from -10 to 8 using a linear scale and the y-axis ranges from the blue dots to the orange dots using a categorical scale. Strip 1 has 350 points, a median of x=0.9834, and 1 outlier at x=-5.01. Strip 2 has 250 points, a median of x=-3.955, and 3 outliers. \\
            \midrule
            \textbf{Heatmap} & A 7x6 heatmap titled 'number of points from combined gaussians'. The x-axis ranges from -8 to 6, the y-axis ranges from -4 to 8, and number of gaussian points is plotted on the z-axis from 0 to 61, all using linear scales. The data has a minimum value of z=0 at (-8, -4), a maximum value of z=61 at (0, -4), and an average of z=14.19.\\
            \midrule
            \textbf{Contour plot} & A contour plot titled 'number of points from combined gaussians'. The x-axis ranges from -12 to 6 and the y-axis ranges from -4 to 8, both using linear scales. 9 contour lines are plotted with values 0, 0, 8, 16, 24, 32, 32, 40, and 48. The max contour is centered around (-4.803, 4.432).\\
            \midrule
            \textbf{Multiple subplots} & A figure with 4 subplots titled 'anscombe's quartet'.
            \textit{subplot 1}: a line plot. The x-axis ranges from 2.5 to 15 and the y-axis ranges from 2 to 12, both using linear scales. The data are plotted in dark blue. The data has a linear fit of y=0.5001x+3, an average of y=7.501, and a standard deviation of y=1.937. The data generally increase up to their max at x=12.
            \textit{subplot 2}: a line plot. The x-axis ranges from 2.5 to 15 and the y-axis ranges from 2 to 10, both using linear scales. The data are plotted in dark blue. The data has a linear fit of y=0.5x+3.001, an average of y=7.501, and a standard deviation of y=1.937. The data strictly increase up to their max at x=11, then strictly decrease.
            \textit{subplot 3}: a line plot. The x-axis ranges from 2.5 to 15 and the y-axis ranges from 4 to 14, both using linear scales. The data are plotted in dark blue. The data has a linear fit of y=0.4997x+3.002, an average of y=7.5, and a standard deviation of y=1.936. The data strictly increase up to their max at x=13, then strictly decrease.
            \textit{subplot 4}: a line plot. The x-axis ranges from 5 to 20 and the y-axis ranges from 4 to 14, both using linear scales. The data are plotted in dark blue. The data has a linear fit of y=0.4999x+3.002, an average of y=7.501, and a standard deviation of y=1.936.\\
            \midrule
            \textbf{Boxplot} & A boxplot titled 'distributional differences in anscombe's quartet'. Subplot num is plotted on the x-axis from 0 to 3 and y value in anscombe's quartet is plotted on the y-axis from 2 to 14, both using linear scales. Boxplot 0 has a median of 7.58, an interquartile range of 2.255, and no outliers. Boxplot 1 has a median of 8.14, an interquartile range of 2.255, and 1 outlier at y=3.1. Boxplot 2 has a median of 7.11, an interquartile range of 1.73, and 1 outlier at y=12.74. Boxplot 3 has a median of 7.04, an interquartile range of 2.02, and 1 outlier at y=12.5.\\
            \midrule
            \textbf{Image} & A 50x50 image titled 'z\^3 + 0.66 iterations before reaching a threshold of 2'. The x-axis ranges from -0.5 to 49.5 using a numerical scale, the y-axis ranges from -0.5 to 49.5 using a linear scale, and iterations is plotted on the z-axis from 0 to 17 using a linear scale. The data has a minimum value of z=0 at (0, 0), a maximum value of z=17 at (38, 12), and an average of z=2.511.\\
            \midrule
            \textbf{Blank plot} & A blank plot titled 'blank'. X is plotted on the x-axis from -0.06 to 0.06 and y is plotted on the y-axis from -0.06 to 0.06, both using linear scales.\\
            \midrule
        \end{tabular}
        \caption{Heuristic figure descriptions for each of the gallery figures created by calling \texttt{show\_with\_alt}}
        \label{tab:example_captions}
    \end{table*}
\end{center}

\begin{table*}[!ht]
    \centering
    \begin{tabular}{|l|l|l|}
    \hline
        month & \makecell{\# bikes crossing fremont bridge\\(change from yearly average (\%))} & \makecell{hours of sunshine\\(change from yearly average (\%))} \\ \hline
        0 & -32.52 & -61.72 \\ \hline
        1 & -37.78 & -40.08 \\ \hline
        2 & -18.13 & -1.248 \\ \hline
        3 & -0.7941 & 14.84 \\ \hline
        4 & 39.35 & 40.36 \\ \hline
        5 & 33.02 & 48.68 \\ \hline
        6 & 40.93 & 73.09 \\ \hline
        7 & 34.72 & 55.89 \\ \hline
        8 & 14.37 & 22.61 \\ \hline
        9 & -0.1582 & -21.22 \\ \hline
        10 & -26.93 & -60.06 \\ \hline
        11 & -46.08 & -71.15 \\ \hline
    \end{tabular}
    \caption{An example markdown table created using \texttt{generate\_alt\_text} (converted to LaTeX formatting)}
    \label{tab:example_md_datatable}
\end{table*}

\definecolor{L1}{HTML}{FFD6B2}
\definecolor{L2}{HTML}{94DE94}
\definecolor{L3}{HTML}{FFFFA8}
\definecolor{L4}{HTML}{2B82D4}

\begin{center}
\begin{table*}
    \resizebox{\textwidth}{!}{%
    \begin{tabular}{M{0.08\linewidth} | M{1.25\columnwidth} | M{1.25\columnwidth}}
        \toprule
        Alt Text Type & Sunshine and Bikes Example Alt Text & Anscombe's Quartet Example Alt Text \\
        \midrule
        \rotatebox{90}{Figure Image} & \includegraphics[width=1.25\columnwidth, alt={See descriptions in the rows below, which have been verified as correct.}]{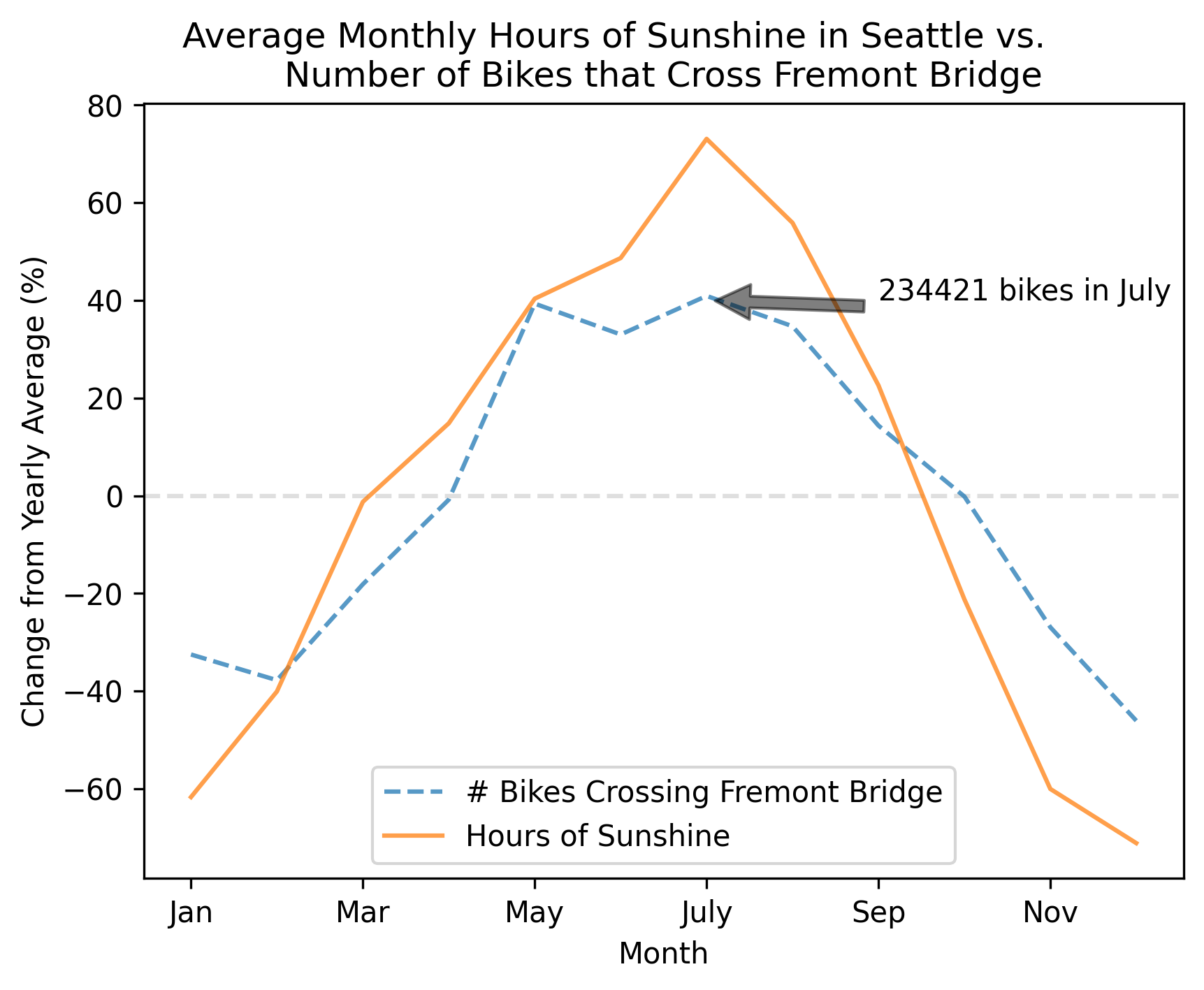} & \includegraphics[width=1.25\columnwidth, alt={See descriptions in the rows below, which have been verified as correct.}]{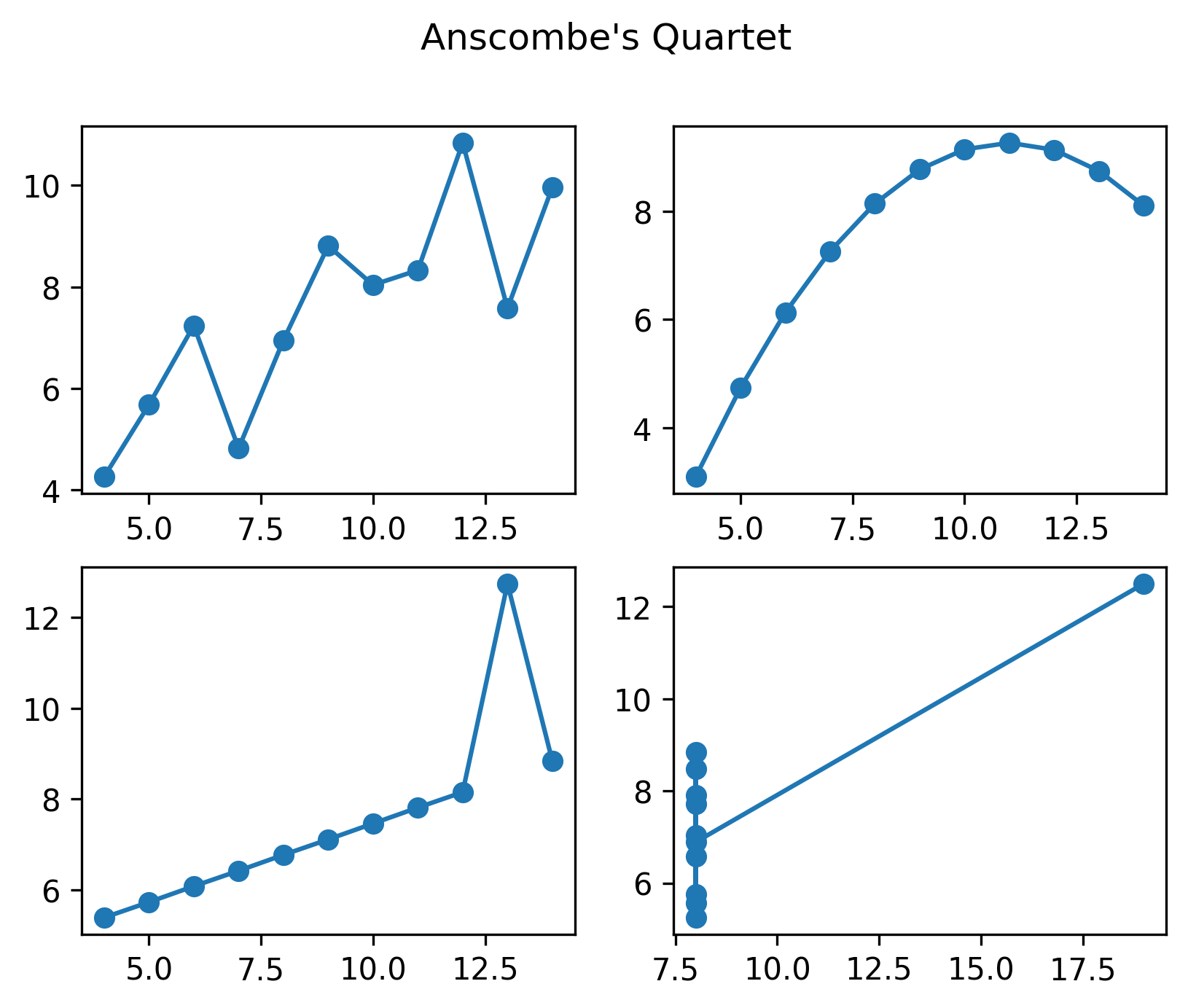} \\
        \midrule
        \rotatebox{90}{Heuristic} &
        \begin{flushleft} [L1] \hlc[L1]{A line plot titled 'average monthly hours of sunshine in seattle vs. Number of bikes that cross fremont bridge'. Month is plotted on the x-axis from jan to nov using a datetime scale and change from yearly average (\%) is plotted on the y-axis from -80 to 100 using a linear scale. \# bikes crossing fremont bridge is plotted in dark blue and hours of sunshine is plotted in orange. There is a horizontal line at y=0.0. An annotation reads '234421 bikes in july'.} [L2] \hlc[L2]{\# bikes crossing fremont bridge has a minimum value of y=-46.08 at x=12, a maximum value of y=40.93 at x=7, and an average of y=0. Hours of sunshine has a minimum value of y=-71.15 at x=12, a maximum value of y=73.09 at x=7, and an average of y=0.} [L3] \hlc[L3]{Hours of sunshine strictly increase up to their max at x=7, then strictly decrease.} [L2] \hlc[L2]{\# bikes crossing fremont bridge and hours of sunshine have a correlation of 0.95.} \end{flushleft} & \scalefont{0.8} \begin{flushleft} [L1] \hlc[L1]{A figure with 4 subplots titled 'anscombe's quartet'. 
        
        \textbf{subplot 1}: a line plot. The x-axis ranges from 2.5 to 15 and the y-axis ranges from 2 to 12, both using linear scales. The data are plotted in dark blue.} [L2] \hlc[L2]{The data has a minimum value of y=4.26 at x=4, a maximum value of y=10.84 at x=12, and an average of y=7.501.} [L3] \hlc[L3]{The data generally increase up to their max at x=12.}
        
        [L1] \hlc[L1]{\textbf{subplot 2}: a line plot. The x-axis ranges from 2.5 to 15 and the y-axis ranges from 2 to 10, both using linear scales. The data are plotted in dark blue.} [L2] \hlc[L2]{The data has a minimum value of y=3.1 at x=4, a maximum value of y=9.26 at x=11, and an average of y=7.501.} [L3] \hlc[L3]{The data strictly increase up to their max at x=11, then strictly decrease.}
        
        [L1] \hlc[L1]{\textbf{subplot 3}: a line plot. The x-axis ranges from 2.5 to 15 and the y-axis ranges from 4 to 14, both using linear scales. The data are plotted in dark blue.} [L2] \hlc[L2]{The data has a minimum value of y=5.39 at x=4, a maximum value of y=12.74 at x=13, and an average of y=7.5.} [L3] \hlc[L3]{The data strictly increase up to their max at x=13, then strictly decrease.}
        
        [L1] \hlc[L1]{\textbf{subplot 4}: a line plot. The x-axis ranges from 5 to 20 and the y-axis ranges from 4 to 14, both using linear scales. The data are plotted in dark blue.} [L2] \hlc[L2]{The data has a minimum value of y=5.25 at x=8, a maximum value of y=12.5 at x=19, and an average of y=7.501.}\end{flushleft}\\
        \midrule
        \rotatebox{90}{turbo} &   \begin{flushleft} [L1] \hlc[L1]{This is a line chart titled 'Average Monthly Hours of Sunshine in Seattle vs. Number of Bikes that Cross Fremont Bridge'. Two variables are plotted: the number of bikes crossing Fremont Bridge is shown with a dashed blue line, and the hours of sunshine with a solid orange line. Both variables are compared against the months of the year, plotted on the horizontal x-axis from January to November. The vertical y-axis represents the change from yearly average, marked in percentages ranging from -60\% to 80\%.} [L2] \hlc[L2]{The number of bikes peaks in July at 234421 bikes, which correlates with the highest point of the hours of sunshine.} [L3] \hlc[L3]{The winter months show a decrease in both bikes crossing and sunshine hours.} [L4] {\color[HTML]{FFFFFF}\hlc[L4]{This suggests a seasonal pattern where more people bike during sunnier months.}} [L2] \hlc[L2]{The chart demonstrates a clear positive correlation between the number of bikers and the amount of sunshine.}\end{flushleft}&   \begin{flushleft}[L1] \hlc[L1]{This figure displays Anscombe's Quartet, consisting of four scatter plots, each depicting a different dataset. The x-axis ranges from about 4 to 20, and the y-axis ranges from roughly 3 to 14 across the four charts.} [L2] \hlc[L2]{All datasets exhibit a similar statistical summary yet present distinct distributions:} 
        
        [L3] \hlc[L3]{\textbf{Top-left plot}: Appears to show a linear relationship with some variability.
        
        \textbf{Top-right plot}: Displays a curved, quadratic-like relationship.
        
        \textbf{Bottom-left plot}: Mostly linear but with a clear outlier that deviates significantly from the trend.
        
        \textbf{Bottom-right plot}: Shows a strong linear relationship with an outlier that lies vertically far from the line.}
        
        [L1] \hlc[L1]{Each plot's data points are in blue with connecting lines.} [L4] {\color[HTML]{FFFFFF}\hlc[L4]{The consistency of these statistical properties, despite differing datasets, underscores the importance of graphically analyzing data to avoid misleading interpretations based only on statistical summaries.}}\end{flushleft}\\
        \midrule
        \rotatebox{90}{turbo + heuristic} &   \begin{flushleft}[L1] \hlc[L1]{This line chart depicts 'Average Monthly Hours of Sunshine in Seattle vs. Number of Bikes that Cross Fremont Bridge'. The x-axis represents months, ranging from January to November, while the y-axis shows the percentage change from the yearly average, ranging from -80\% to +100\%. Two data series are plotted: the number of bikes crossing Fremont Bridge is illustrated with a dark blue dashed line, and the hours of sunshine are shown with a solid orange line. A horizontal reference line is drawn at y=0\%.}

        [L2] \hlc[L2]{The data shows that both variables reach their peak in July, with the hours of sunshine increasing to a maximum of 73.09\% and the number of bikes reaching a peak change of 40.93\%, marked by an annotation indicating "234421 bikes in July". Conversely, December experiences the lowest points for both variables, with bikes at -46.08\% and sunshine at -71.15\%. The overall trend suggests a strong correlation (0.95) between sunshine and cycling activity,} [L4] {\color[HTML]{FFFFFF}\hlc[L4]{reflecting a potential increase in outdoor activities during sunnier months.}}\end{flushleft}&   \begin{flushleft}[L1] \hlc[L1]{The figure consists of four subplots collectively titled "Anscombe's Quartet," each displaying a line plot with data points in dark blue.

        \textbf{Subplot 1}: x-axis 2.5-15, y-axis 2-12.} [L2/L3] \hlc[L3]{Data rises to y=10.84 at x=12 then falls slightly.}

        \textbf{Subplot 2}: [L1] \hlc[L1]{x-axis 2.5-15, y-axis 2-10.} [L2/L3] \hlc[L3]{Data climbs to y=9.26 at x=11 then decreases.}
        
        \textbf{Subplot 3}: [L1] \hlc[L1]{x-axis 2.5-15, y-axis 4-14.} [L2/L3] \hlc[L3]{Data ascends to y=12.74 at x=13, followed by a decline.}
        
        \textbf{Subplot 4}: [L1] \hlc[L1]{x-axis 5-20, y-axis 4-14.} [L2/L3] \hlc[L3]{Data shows a strong linear relationship from the lower left to the upper right corner.}
        
        [L2] \hlc[L2]{Each subplot shares a mean of y=7.5 (approx.)} [L4] {\color[HTML]{FFFFFF}\hlc[L4]{but displays distinct patterns, illustrating how identical statistical properties can arise from very different data distributions.}}\end{flushleft}\\
        \bottomrule
    \end{tabular}}
    \caption{MatplotAlt example alt text for two figures using each generation strategy. Continued in Table \ref{table:example_descriptions2} for turbo + table and turbo + table + heuristic methods. Text is color-coded and tagged by semantic level. Emphasis Added for readability.}
    \label{table:example_descriptions1}
\end{table*}
\end{center}

\begin{center}
\begin{table*}
    \resizebox{\textwidth}{!}{%
    \begin{tabular}{M{0.08\linewidth} | M{1.25\columnwidth} | M{1.25\columnwidth}}
        \toprule
        Alt Text Type & Sunshine and Bikes Example Alt Text & Anscombe's Quartet Example Alt Text \\
        \toprule
        \rotatebox{90}{turbo + heuristic + table} & \begin{flushleft}[L1]\hlc[L1]{This is a line chart titled 'Average Monthly Hours of Sunshine in Seattle vs. Number of Bikes that Cross Fremont Bridge'. It compares the monthly change from the yearly average for the number of bikes crossing Fremont Bridge (dashed line) and the hours of sunshine (solid line) in Seattle. The horizontal x-axis labels the months from January to December, while the vertical y-axis represents the percent change from the yearly average, ranging from -80\% to 60\%.}\\~\\~\\~
        [L3] \hlc[L3]{Both variables show seasonality with values rising in the warmer months and falling in the cooler months.} [L2] \hlc[L2]{The highest point is in July for both variables, with hours of sunshine around 73\% above the average and a peak in bike crossings at 234421, which corresponds to approximately a 41\% increase. The chart suggests a positive correlation between the amount of sunshine and the number of bicycles crossing the bridge.}\end{flushleft} & \begin{flushleft}[L1] \hlc[L1]{This figure presents Anscombe's Quartet, comprising four subplots of line plots, each depicting a different trend. The title "Anscombe's Quartet" is located at the top. Each plot has its data points marked in dark blue}, [L3] \hlc[L3]{with the plots varying in data distribution and trends.}

        [L1] \hlc[L1]{\textbf{Subplot 1}: The x-axis ranges from 4 to 14, and the y-axis ranges from 2 to 12.} [L3] \hlc[L3]{The data display a somewhat erratic pattern, with fluctuations but generally trending upwards.} [L2] \hlc[L2]{A peak is observed at y=10.84 with x=12.}
        
        [L1] \hlc[L1]{\textbf{Subplot 2}: The x-axis is from 4 to 14, and the y-axis from 2 to 10.} [L2 / L3] \hlc[L3]{The data exhibit a highly curved trend, first rising to a peak of y=9.26 at x=11 then declining smoothly.}
        
        [L1] \hlc[L1]{\textbf{Subplot 3}: The x-axis is between 4 to 14, and the y-axis extends from 4 to 14.} [L2 / L3] \hlc[L3]{The line plot shows a continuous increase up to a sharp peak at y=12.74, x=13, followed by a drop.}
        
        [L1] \hlc[L1]{\textbf{Subplot 4}: Here, the x-axis ranges from 8 to 19, while the y-axis spans from 4 to 14.} [L3] \hlc[L3]{The data is tightly clustered at x=8, except for an outlier at x=19} \end{flushleft} \\
        \midrule
        \rotatebox{90}{turbo + heuristic + table} &   \begin{flushleft}[L1] \hlc[L1]{This is a line chart titled 'Average Monthly Hours of Sunshine in Seattle vs. Number of Bikes That Cross Fremont Bridge'. The months are plotted from January to November on the horizontal x-axis and the change from yearly average is denoted in percentage on the vertical y-axis ranging from -80\% to 100\%. The number of bikes crossing Fremont Bridge is represented by a dashed line in dark blue, while hours of sunshine is illustrated with a solid orange line.} [L2] \hlc[L2]{There is a strong positive correlation between the two variables, with a correlation coefficient of 0.95. Both metrics peak in July, where the number of bikes reaches a 40.93\% increase from the yearly average, with 234,421 bikes recorded in that month. Simultaneously, hours of sunshine cap at a 73.09\% increase.} [L3] \hlc[L3]{Both variables exhibit similar seasonal patterns, with higher values in the warmer months and lower values in the colder months,} [L4] {\color[HTML]{FFFFFF}\hlc[L4]{reflecting the influence of weather on outdoor activities such as biking.}}\end{flushleft}&   \begin{flushleft}[L1] \hlc[L1]{This figure, titled 'Anscombe's Quartet', consists of four subplots with line plots of dark blue data points.} [L4] {\color[HTML]{FFFFFF}\hlc[L4]{Each chart has a distinct pattern, despite having the same mean and variance, demonstrating different data distributions that yield similar descriptive statistics.}}

        [L1] \hlc[L1]{\textbf{Subplot 1}'s x-axis ranges from 2.5 to 15, and y-axis from 2 to 12.} [L2/L3] \hlc[L3]{Data points generally increase, reaching a maximum at x=12, y=10.84.}

        [L1] \hlc[L1]{\textbf{Subplot 2}'s x- and y-axes share the same range as the first,} [L2/L3] \hlc[L3]{but the data strictly increase to a maximum at x=11, y=9.26, then strictly decrease.}

        [L1] \hlc[L1]{\textbf{Subplot 3}'s x-axis is the same, with the y-axis ranging from 4 to 14.} [L2/L3] \hlc[L3]{Data strictly increase, peaking at x=13, y=12.74, before decreasing.}

        [L1] \hlc[L1]{\textbf{Subplot 4} features an x-axis from 5 to 20 and a y-axis from 4 to 14.} [L2/L3] \hlc[L3]{Most data cluster at x=8, with an outlier at x=19, y=12.5, creating a steep incline.}

        [L4] {\color[HTML]{FFFFFF}\hlc[L4]{This quartet is commonly used to illustrate the importance of graphing data before analyzing it and the limitations of summary statistics.}}\end{flushleft}\\
        \bottomrule
    \end{tabular}}
    \caption{MatplotAlt example alt text for two figures using each generation strategy. Continued in Table \ref{table:example_descriptions1} for Heuristic, turbo, and turbo + heuristic methods. Text is color-coded and tagged by semantic level. Emphasis Added for readability.}
    \label{table:example_descriptions2}
\end{table*}
\end{center}

\definecolor{aquamarine}{HTML}{7FFFD4}
\definecolor{lightorange}{HTML}{ffbf00}
\definecolor{lightpurple}{HTML}{df73ff}
\definecolor{wheat}{HTML}{f5deb3} 
 
\begin{table*}
    \resizebox{\textwidth}{!}{%
    \begin{tabular}{M{1.35\columnwidth} | M{1.35\columnwidth}}
        \toprule
        VisText Example Alt Text & Matplotlib Gallery Example Alt Text \\
        \midrule
        \includegraphics[width=1.25\columnwidth, alt={An area chart titled "Number of shopping malls inthe United States with less than 100,001 sq. feet of gross leasable area from 1986 to 2010". Number of shopping malls is plotted on the y-axis from 0 to 100000, and year is plotted on the x-axis from 1985 to 2010. The data starts at it's min slightly below 60000 malls, then steadily increases up to it's max in 2009. There is no data for the year 2010, but the increase in malls appears to plateau at the end of the time period shown.}]{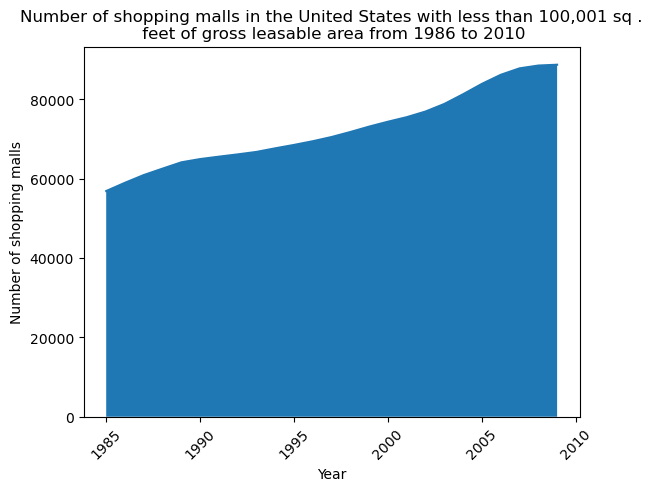} & \includegraphics[width=1.25\columnwidth, alt={A contour plot titled 100^Z. Both axes are unlabeled and there are no variable names. The y-axis ranges from -2 to 2, and the x-axis ranges from -3 to 3, both using linear scales. The contours range from 10^3 (yellow) centered near the origin to 10^-3 (dark purple) centered around approximately (1.25, 1.25). There is a diagonal contour extending from the top left/middle to the bottom of the right of the plot that divides these regions of high and low value.}]{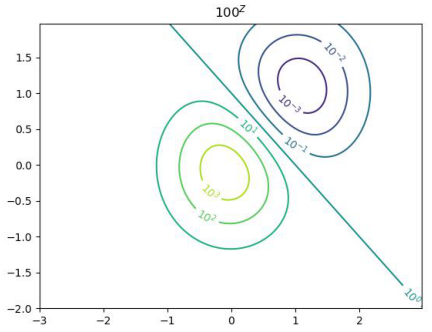} \\
        \midrule
        \begin{flushleft}\textbf{turbo description:} \hlc[lime]{This is an area chart titled "Malls in the United States with less than 100,001 sq. feet of gross leasable area." [\textit{Label error}]} It plots the number of shopping malls on the vertical y-axis, which ranges from 0 to 80,000 in increments of 20,000. The years are plotted on the horizontal x-axis from 1985 to around 2010. The chart shows a noticeable increase in the number of shopping malls over time, starting at about \hlc[lightorange]{20,000 malls in 1985 and rising to just under 80,000 by 2010 [\textit{Value error}]}. There's a steady growth up to around 1995, \hlc[cyan]{after which the rate of increase becomes more gradual, reaching a plateau in the mid-2000s before slightly decreasing towards 2010. [\textit{Trend error}]}\end{flushleft} & \begin{flushleft}\textbf{Heuristic description:} A contour plot titled '$100^z$'. The x-axis ranges from -3 to 3 and the y-axis ranges from -2 to 2, both using linear scales. 7 contour lines are plotted with values ${10^{-3}}$, ${10^{-2}}$, ${10^{-1}}$, ${10^{0}}$, ${10^{1}}$, ${10^{2}}$, and ${10^{3}}$. The max contour is centered around (-0.06283, -0.05828). \hlc[pink]{[\textit{Missing data context:} lacks description of contour shape including the diagonal line and minimum]}\end{flushleft}\\
        
        \midrule
        \begin{flushleft}\textbf{turbo + table description:} This is an area chart showing the number of shopping malls in the United States with less than 100,001 square feet of gross leasable area over time. The horizontal x-axis ranges from the year 1985 to 2010, while the vertical y-axis represents the number of shopping malls from 0 to 80,000. The chart shows a substantial increase in the number of small malls from 1985, \hlc[yellow]{peaking between 2000 and 2005 [\textit{Identity error}]}, and then a plateau or slight decrease towards 2010. There are no clear outliers, and the trend shows overall growth in mall numbers over time, with a stabilization or minor decline in the latter part of the series. \hlc[pink]{[\textit{Missing data context:} lacks chart values, e.g. number of malls in 1985]} \end{flushleft} & \begin{flushleft}\textbf{turbo description:} This is a contour plot with labels indicating different levels, likely representing a mathematical function's value over a two-dimensional space. The horizontal x-axis ranges from -3 to 3, while the vertical y-axis ranges from -2 to 2. \hlc[aquamarine]{Contour lines are presented with four distinct levels, indicated by 10\textasciicircum-2, 10\textasciicircum-1, 10\textasciicircum0, and 10\textasciicircum1 [\textit{Axis error}: z-axis contours extend from 10\textasciicircum-3 to 10\textasciicircum3]}, each represented in a different color from yellow (innermost) to dark blue (outermost). A cyan diagonal line, annotated as 10\textasciicircum2, suggests an additional function or constraint cutting across the plot. \hlc[lightpurple]{The title is partially visible, reading only "100z." The plot shows a symmetrical pattern about the origin, with each contour line indicating a constant value. [\textit{Deceptive error}: title is fully visible and pattern is symmetric around the diagonal line, not the origin]}\end{flushleft}\\

        \midrule
        \begin{flushleft}\textbf{turbo + heuristic description:} This is an area chart titled 'Number of shopping malls in the United States with less than 100,001 sq. feet of gross leasable area from 1986 to 2010'. The year is plotted on the horizontal x-axis from 1980 to 2015, and the number of shopping malls is plotted on the vertical y-axis from 0 to 100,000, both using linear scales. The chart shows one variable plotted over time. The data exhibits a strict increasing trend from the year 1985, where the number of shopping malls is at its minimum of 56,910, peaking at its maximum in the year 2009 with 88,760 shopping malls. There are no apparent outliers, and the strict increase indicates a consistent growth in the number of shopping malls with fewer than 100,001 sq. feet of leasable area over the given time period. After reaching the peak in 2009, there seems to be a slight plateau or very gradual decline towards 2010, though the data does not extend beyond that year to visualize any further trends.\\\hlc[green]{[We consider this alt text correct]}\end{flushleft} & \begin{flushleft}\textbf{turbo + heuristic description:} This contour plot is titled '$100^z$' and visually represents three-dimensional data on a two-dimensional plane. The x-axis is labeled and ranges from -3 to 3, while the y-axis ranges from -2 to 2. There are seven contour lines corresponding to values ${10^{-3}}$, ${10^{-2}}$, ${10^{-1}}$, ${10^{0}}$, ${10^{1}}$, ${10^{2}}$, and ${10^{3}}$. These lines indicate the points on the plane where the function $100^z$ has the same value. The contour values are increasing as they move towards the center, with the max contour centered approximately at (-0.06283, -0.05828). The colors transition from green (lower values) to yellow (higher values) \hlc[lightpurple]{and then to blue for the highest values towards the center of the plot. [\textit{Deceptive error:} Blue/purple colors represent lowest values]} The plot lacks a legend, but the contour values are labeled directly on the lines. The overall trend shows that \hlc[wheat]{[\textit{Cutoff}]} \end{flushleft}\\
        \bottomrule
    \end{tabular}}
    \caption{Example descriptions containing errors for one VisText figure, and one Matplotlib gallery figure. Errors are colored and tagged by type.}
    \label{table:example_error_descriptions}
\end{table*}

%% file: supp_content/templates.tex
\section{Heuristic Alt Text Templates}
\label{sec:heuristic_templates}

While our heuristic alt text is formatted differently from figure to figure, we use the following templates in most cases. Our L1 description is usually:
\begin{displayquote}
    \texttt{a [chart type] titled [title]. [x-label] is plotted on the x-axis from [x-axis min] to [x-axis max] using a [x-axis scale] scale. [y-label] is plotted on the y-axis from [y-axis min] to [y-axis max], ...}
\end{displayquote}
For each variable encoded with a color we add:
\begin{displayquote}
    \texttt{[variable label] is plotted in [label color encoding].}
\end{displayquote}
And for each annotation we add:
\begin{displayquote}
    \texttt{An annotation at [annotation x/y position] reads [annotation text].}
\end{displayquote}
For L2 descriptions, we compute relevant statistics from the data depending on the chart type and then describe them as:
\begin{displayquote}
    \texttt{[variable label] has a [statistic name] of [dependent axis]=[value] at [independent axis]=[value], a [statistic name] of....}
\end{displayquote}
Several statistics do not fit this format. For example: \texttt{The max contour is centered around [x, y]}.

For plots displaying one or two-dimensional data, we generate simple trend summaries from the difference between subsequent points. For instance:
\begin{displayquote}
    \texttt{[variable name] strictly increase to a max at [dependent axis]=[end point], then generally decrease.}
\end{displayquote}
Future implementations may use more involved methods to capture complex trends like fitting piecewise functions. To see a range of the different outputs possible with \texttt{generate\_alt\_text}, we provide more examples at \URL{https://github.com/make4all/matplotalt/blob/main/examples/examples.ipynb}

%% file: supp_content/vlm_prompts.tex
\section{Vision Language Model Prompts}
\label{sec:vlm_prompts}

We use different versions of our instruction prompt based on the given description level and other config options.

\begin{center}\textbf{L1:}\end{center}~\\ \texttt{You are a helpful assistant that describes figures. Here are two example descriptions:
1. 'This is a vertical bar chart entitled 'COVID-19 mortality rate by age' that plots Mortality rate by Age. Mortality rate is plotted on the vertical y-axis from 0 to 15\%. Age is plotted on the horizontal x-axis in bins: 10-19, 20-29, 30-39, 40-49, 50-59, 60-69, 70-79, 80+.'
2. 'This is a line chart titled 'Big Tech Stock Prices' that plots price by date. The corporations include AAPL (Apple), AMZN (Amazon), GOOG (Google), IBM (IBM), and MSFT (Microsoft). The years are plotted on the horizontal x-axis from 2000 to 2010 with an increment of 2 years. The prices are plotted on the vertical y-axis from 0 to 800 with an increment of 200.'}

\texttt{Only include information about the chart type, title, axis ranges, and labels. Be concise and limit your response to \{max\_tokens\} tokens.}

\begin{center}\textbf{L2:}\end{center}~\\ \texttt{You are a helpful assistant that describes figures. Here are two example descriptions:
1. 'This is a vertical bar chart entitled 'COVID-19 mortality rate by age' that plots Mortality rate by Age. Mortality rate is plotted on the vertical y-axis from 0 to 15\%. Age is plotted on the horizontal x-axis in bins: 10-19, 20-29, 30-39, 40-49, 50-59, 60-69, 70-79, 80+. The highest COVID-19 mortality rate is in the 80+ age range, while the lowest mortality rate is in 10-19, 20-29, 30-39, sharing the same rate. COVID-19 mortality rate does not linearly correspond to the demographic age.'
2. 'This is a line chart titled 'Big Tech Stock Prices' that plots price by date. The corporations include AAPL (Apple), AMZN (Amazon), GOOG (Google), IBM (IBM), and MSFT (Microsoft). The years are plotted on the horizontal x-axis from 2000 to 2010 with an increment of 2 years. The prices are plotted on the vertical y-axis from 0 to 800 with an increment of 200. GOOG has the greatest price over time. MSFT has the lowest price over time.'}

\texttt{Include information about the chart type, title, axis ranges, and labels. If possible, describe statistics, extrema, outliers, correlations, and point-wise comparisons between variables. Be concise and limit your response to \{max\_tokens\} tokens.}

\begin{center}\textbf{L3:}\end{center}~\\ \texttt{You are a helpful assistant that describes figures. Here are two example descriptions:
1. 'This is a vertical bar chart entitled 'COVID-19 mortality rate by age' that plots Mortality rate by Age. Mortality rate is plotted on the vertical y-axis from 0 to 15\%. Age is plotted on the horizontal x-axis in bins: 10-19, 20-29, 30-39, 40-49, 50-59, 60-69, 70-79, 80+. The highest COVID-19 mortality rate is in the 80+ age range, while the lowest mortality rate is in 10-19, 20-29, 30-39, sharing the same rate. COVID-19 mortality rate does not linearly correspond to the demographic age. The mortality rate increases with age, especially around 40-49 years and upwards. The mortality rate increases exponentially with older people.'
2. 'This is a line chart titled 'Big Tech Stock Prices' that plots price by date. The corporations include AAPL (Apple), AMZN (Amazon), GOOG (Google), IBM (IBM), and MSFT (Microsoft). The years are plotted on the horizontal x-axis from 2000 to 2010 with an increment of 2 years. The prices are plotted on the vertical y-axis from 0 to 800 with an increment of 200. GOOG has the greatest price over time. MSFT has the lowest price over time. Prices of particular Big Tech corporations seem to fluctuate but nevertheless increase over time. Years 2008-2009 are exceptions as we can see an extreme drop in prices of all given corporations.'}

\texttt{Include information about the chart type, title, axis ranges, and labels. If possible, describe statistics, extrema, outliers, correlations, point-wise comparisons, and trends for each plotted variable. Be concise and limit your response to \{max\_tokens\} tokens.}

\begin{center}\textbf{L4:}\end{center}~\\ \texttt{You are a helpful assistant that describes figures. Here are two example descriptions:
1. 'This is a vertical bar chart entitled 'COVID-19 mortality rate by age' that plots Mortality rate by Age. Mortality rate is plotted on the vertical y-axis from 0 to 15\%. Age is plotted on the horizontal x-axis in bins: 10-19, 20-29, 30-39, 40-49, 50-59, 60-69, 70-79, 80+. The highest COVID-19 mortality rate is in the 80+ age range, while the lowest mortality rate is in 10-19, 20-29, 30-39, sharing the same rate. COVID-19 mortality rate does not linearly correspond to the demographic age. The mortality rate increases with age, especially around 40-49 years and upwards. This relates to people’s decrease in their immunity and the increase of co-morbidity with age. The mortality rate increases exponentially with older people.'
2. 'This is a line chart titled 'Big Tech Stock Prices' that plots price by date. The corporations include AAPL (Apple), AMZN (Amazon), GOOG (Google), IBM (IBM), and MSFT (Microsoft). The years are plotted on the horizontal x-axis from 2000 to 2010 with an increment of 2 years. The prices are plotted on the vertical y-axis from 0 to 800 with an increment of 200. GOOG has the greatest price over time. MSFT has the lowest price over time. Prices of particular Big Tech corporations seem to fluctuate but nevertheless increase over time. Years 2008-2009 are exceptions as we can see an extreme drop in prices of all given corporations. The big drop in prices was caused by financial crisis of 2007-2008. The crisis culminated with the bankruptcy of Lehman Brothers on September 15, 2008 and an international banking crisis.'}

\texttt{Include information about the chart type, title, axis ranges, and labels. If possible, describe statistics, extrema, outliers, correlations, point-wise comparisons, and trends for each plotted variable. If possible, briefly explain domain-specific insights, current events, and socio-political context that explain the data. Be concise and limit your response to \{max\_tokens\} tokens.}

To incorporate heuristic alt text and markdown data tables, we include the following line between the example descriptions and instructions about description content: \texttt{You already know the following information about this figure and its data: '\{generated heuristic alt text, possibly containing a table\}'.}

By default, we also prompt models to describe \texttt{colors, sizes, textures}, but removed this clause for our evaluation to align with crowdsourced VisText captions.

%% file: supp_content/ngram_similarity.tex
\section{N-gram similarity between generated and reference captions}
\label{sec:ngram_similarity}

In this section we measure BLEU, Rouge, and CHRF n-gram similarity between our generated and reference captions for VisText figures. Table \ref{table:ngram_similarity_scores} summarizes our results. We find that VL-T5 models typically have higher n-gram overlap with the human reference than  GPT4-turbo, while our turbo + heuristic methods have higher Rouge-1 and Rouge-L recalls. This indicates turbo + heuristic descriptions contain overall more tokens from crowdsourced text compared to turbo and turbo + table methods, but do not emulate the style and formatting of the references as well as the finetuned VL-T5 models.

\begin{table*}[ht!]
\resizebox{\textwidth}{!}{%
\begin{tabular}{l | *{13}{M{0.065\linewidth}}}
    \toprule
    VisText Alt Text Type       & BLEU                                   & Rouge-1 F1                             & Rouge-1 recall                         & Rouge-2 F1                             & Rouge-2 recall                         & Rouge-L F1                             & Rouge-L recall                         & CHRF                                  \\ \hline
VL-T5 - image only          & 5.504          & 33.28          & 30.91          & 13.01          & 12.14          & 24.09          & 22.43          &   26.36          \\
VL-T5 - image + scene graph & \textbf{29.47} & \textbf{59.83} & 52.03          & \textbf{43.55} & \textbf{37.88} & \textbf{49.29} & 42.90          & \textbf{50.59}           \\
VL-T5 - image + table       & \textbf{29.45} & \textbf{60.30} & 51.90          & \textbf{43.93} & \textbf{37.83} & \textbf{49.75} & 42.86          & \textbf{50.89}          \\ \hline
Heuristic                   & 15.56          & 54.60          & 58.37          & 32.51          & 35.04          & 39.56          & 42.50          & 45.89          \\
turbo                       & 14.14          & 46.37          & 63.77          & 22.61          & 31.47          & 32.32          & 44.77          & 46.45 \\
turbo + table               & 13.46          & 46.06          & 66.00          & 22.35          & 32.47          & 31.97          & 46.23          & 46.62 \\
turbo + heuristic           & 15.66          & 48.37          & 67.11          & 25.97          & 36.52          & 35.00          & \textbf{48.96} & 49.22 \\
turbo + table + heuristic   & 15.02          & 47.55          & \textbf{67.68} & 25.50          & 36.84          & 34.32          & \textbf{49.26} & 48.94 \\ \hline
\end{tabular}}
    \caption{N-gram similarity scores between crowdsourced and generated descriptions for each method. Higher is better for each metric. VisText VL-T5 models typically have higher n-gram overlap with the human references, while turbo + heuristic generations have higher Rouge-2 and Rouge-L recalls.}
    \label{table:ngram_similarity_scores}
\end{table*}

%% file: supp_content/shuffling.tex
\section{Similarity between shuffled images and descriptions}
\label{sec:shuffling_experiments}
To measure whether similarity metrics from our evaluation are biased towards specific generation strategies, we compute each score using random pairs of alt text and references. For our comparisons to crowdsourced captions in \S5.2, we randomly sample references for each MatplotAlt and VL-T5 description. For our measurement of similarity between alt texts and images using BLIP in \S5.3, we select corresponding images randomly. A higher score in this setting indicates that the metric is biased towards text from that generation strategy regardless of  whether its content is aligned to the figure. 

We report all shuffled scores between VisText descriptions and crowdsourced references in Table \ref{table:shuffled_similarity_scores}.
All token n-gram overlap F1 scores decreased substantially in this setting except for VL-T5 - imageonly, suggesting that these similarity metrics are not biased towards any one generation method. However, CHRF, Rouge-1 recall, BertScore recall, and BLEURT are all consistently higher for MatplotAlt methods, possibly due to their longer generations and the higher fluency of GPT4 generations.

Our shuffled BLIP results are in Table \ref{table:shuffled_blip_scores}. Matching figures and descriptions were overwhelmingly scored higher than the random pairs, with an average difference of 0.60 in probability and 0.12 in cosine similarity. On the gallery dataset, we find that the shuffled turbo captions scored the worst, indicating that its high score on the matching pairs is not due to a preference for GPT4-generated texts.

\begin{table*}
    \resizebox{\textwidth}{!}{%
    \begin{tabular}{l | *{13}{M{0.065\linewidth}}}
        \toprule
        VisText Alt Text Type & BLEU & Rouge-1 F1 & Rouge-1 recall & Rouge-2 F1 & Rouge-2 recall & Rouge-L F1 & Rouge-L recall & CHRF & BertScore F1 & BertScore recall & BLEURT \\
        \midrule
        VL-T5 - image only & \textbf{4.372} & 31.29 & 29.23 & 11.33 & \textbf{10.64} & \textbf{22.90} & 21.45 & 25.09 & \textbf{0.8565} & 0.8499 & -1.084\\
        VL-T5 - image + scene graph & 4.059 & 29.90 & 26.08 & \textbf{11.57} & 10.12 & 21.54 & 18.83 & 23.55 & \textbf{0.8559} & 0.8458 & -1.009\\
        VL-T5 - image + table & 4.063 & 30.05 & 25.89 & \textbf{11.88} & 10.26 & 21.75 & 18.76 & 23.76 & \textbf{0.8564} & 0.8458 & -1.007\\
        \midrule
        Heuristic & 3.109 & \textbf{34.51} & 36.77 & 10.35 & \textbf{11.05} & 21.47 & 22.99 & 27.86 & 0.8458 & \textbf{0.8520} & -0.8991 \\
        turbo & 2.170 & 28.48 & 39.18 & 5.909 & 8.227 & 17.78 & 24.62 & 28.86 & 0.8440 & \textbf{0.8522} & \textbf{-0.7965} \\
        turbo + table & 2.116 & 28.13 & 40.24 & 5.868 & 8.504 & 17.53 & 25.27 & 28.80 & 0.8424 & \textbf{0.8522} & \textbf{-0.7944} \\
        turbo + heuristic & 2.621 & 29.64 & 40.89 & 7.541 & 10.53 & 18.59 & \textbf{25.85} & \textbf{29.88} & 0.8455 & \textbf{0.8551} & -0.8234 \\
        turbo + table + heuristic & 2.510 & 29.23 & \textbf{41.46} & 7.419 & \textbf{10.63} & 18.29 & \textbf{26.13} & \textbf{29.79} & 0.8443 & \textbf{0.8549} & -0.8291 \\
        \bottomrule
    \end{tabular}}
    \caption{Similarity scores for non-matching shuffled pairs of VisText crowdsourced captions and MatplotAlt-generated descriptions.}
    \label{table:shuffled_similarity_scores}
\end{table*}

\begin{table}[]
    \centering
    \resizebox{\columnwidth}{!}{%
    \begin{tabular}{P{0.25\linewidth} | p{0.35\linewidth} | P{0.25\linewidth} | P{0.25\linewidth}}
        \toprule
        Dataset & Alt Text Type & Matching Prob. & Cosine Sim. \\
        \toprule
        VisText & Human & 0.9999 / 0.3166 & 0.4992 / 0.3570\\
        VisText & Heuristic & 0.9999 / 0.3127 & 0.5037 / 0.3576\\
        VisText & turbo & 0.9985 / 0.2653 & 0.5039 / 0.3675\\
        VisText & turbo + heuristic & 0.9999 / 0.3282 & 0.5079 / 0.3677\\
        VisText & turbo + table & 0.9977 / 0.2726 & 0.5050 / 0.3697\\
        VisText & turbo + heuristic + table & 0.9999 / 0.3170 & 0.5081 / 0.3678\\
        \midrule
        Gallery & Heuristic & 0.9270 / 0.5925 & 0.4263 / 0.3652\\
        Gallery & turbo & \textbf{0.9766} / 0.4897 & \textbf{0.4879} / 0.3883\\
        Gallery & turbo + heuristic & 0.9629 / 0.5241 & 0.4717 / 0.3816\\
        \bottomrule
    \end{tabular}}\\
    \caption{BLIP scores for matching / shuffled pairs of MatplotAlt descriptions and images from each dataset.}
    \label{table:shuffled_blip_scores}
\end{table}

%% file: supp_content/correlations.tex
\section{Correlation Between Evaluation Metrics and Errors}
\label{sec:correlations}

We measure correlations between each observed error type and evaluation metrics, including description length, n-gram and BERT-based similarity to human references, and BLIP scores between images and captions. Table \ref{table:correlations} summarizes our results averaged over both VisText and Matplotlib Gallery datasets and each of our Heuristic and turbo methods. 

On one hand, we find that several of the error types have no significant (p > 0.005) correlations with our other evaluation metrics including axis, value, trend, number name, and nonsense. This suggests that even captions with high similarity to a human-written reference or image embedding may contain these errors. On the other, several error types are correlated to many of the other metrics, including label, cutoff, and deceptive. Description length is also  correlated to several of the caption similarity metrics, particularly their precision components. This is not surprising as longer descriptions likely contain words and meanings not present in the shorter human references, and confirms that the n-gram matching metrics are likely not as relevant due to this discrepancy in lengths. At the same time, Rouge2 F1 has the highest correlation to captions being labeled as correct or value-correct, suggesting that length and word overlap still play a significant role in our manual evaluation of correctness.

\begin{table*}[]
\resizebox{\textwidth}{!}{%
\begin{tabular}{l|c|cc|cccccccccccccc}
\toprule
\begin{tabular}[c]{@{}c@{}}Len. \& error type $\rightarrow$ \\ Correlated metric $\downarrow$ \end{tabular} & \begin{tabular}[c]{@{}c@{}}Description\\length\end{tabular} & Correct & \begin{tabular}[c]{@{}l@{}}Value-\\ correct\end{tabular} & CTE     & AE      & VE      & IE      & TE      & LE      & MC      & UC      & R       & C       & NS      & DE      & NE      \\ \hline
Description len. & N/A                                                                     & $-$0.131  & $-$0.110                                                   & - & -  & -  & -  & - & 0.150 & $-$0.161  & - & \textbf{0.193}   & \textbf{0.277}   & -  & 0.201   & - \\
BLEU & $-$0.233                                                                                  & 0.191   & 0.162                                                    & \textbf{$-$0.206}  & - & -  & - & - & - & - & -  & - & - & - & - & -  \\
Rouge1 (f1) & $-$0.523                                                                            & 0.165   & -                                                  & - & - & - & -  & - & $-$0.174  & - & - & -  & $-$0.148  & - & $-$0.128  & - \\
Rouge1 (p) & $-$0.660                                                                             & 0.133   & -                                                  & -  & - & - & -  & - & $-$0.191  & - & - & -  & $-$0.175  & - & $-$0.151  & - \\
Rouge1 (r) & 0.397                                                                             & -  & -                                                  & - & - & - & -  & - & -  & - & - & - & 0.138   & -  & -  & -  \\
Rouge2 (f1) & $-$0.534                                                                           & \textbf{0.242}   & \textbf{0.218}                                                    & - & - & - & - & - & $-$0.216  & - & - & -  & - & - & \textbf{$-$0.220}  & - \\
Rouge2 (p) & $-$0.688                                                                             & 0.211   & 0.188                                                    & - & -  & - & - & - & $-$0.233  & - & - & -  & $-$0.132  & - & \textbf{$-$0.221}  & - \\
Rouge2 (r) & -                                                                            & 0.167   & 0.164                                                    & - & - & - & - & - & - & $-$0.131  & -  & - & -  & -  & - & - \\
RougeL (f1) & $-$0.535                                                                            & 0.200   & 0.178                                                    & - & - & - & - & - & \textbf{$-$0.269}  & - & - & -  & $-$0.144  & - & \textbf{$-$0.219} & - \\
RougeL (p) & \textbf{$-$0.713}                                                                             & 0.166   & 0.142                                                    & -  & - & - & - & - & $-$0.257  & - & - & -  & $-$0.182  & - & $-$0.211  & - \\
RougeL (r) & 0.241                                                                             & -  & -                                                   & - & -  & - & - & - & - & - & \textbf{0.144}   & - & -  & - & - & -  \\
CHRF & 0.052                                                                                  & 0.193   & 0.136                                                    & $-$0.155  & - & -  & - & - & - & $-$0.207  & -  & - & - & - & - & - \\
BertScore (f1) & $-$0.298                                                                        & 0.144   & -                                                  & - & - & - & - & - & $-$0.207  & - & - & - & $-$0.192  & - & $-$0.158  & -  \\
BertScore (p) & $-$0.428                                                                          & - & -                                                 & - & -  & - & - & - & $-$0.188  & - & - & - & $-$0.262  & - & $-$0.138  & - \\
BertScore (r) & -                                                                        & 0.127   & -                                                   & - & - & -  & - & - & $-$0.129  & - & - & - & - & - & - & - \\
BLEURT & $-$0.043                                                                                & -  & -                                                   & -  & - & -  & -  & -  & $-$0.142  & - & -  & - & - & - & - & -  \\
BLIP matching prob. & -                                                                   & 0.126   & -                                                 & -  & - & -  & - & -  & -  & $-$0.140  & - & - & $-$0.117  & -  & -  & - \\
BLIP cosine sim. & -                                                                     & \textbf{0.237}   & -                                                  & 0.110   & - & - & \textbf{$-$0.166}  & -  & - & \textbf{$-$0.239}  & -  & $-$0.147  & $-$0.126  & - & -  & - \\
\bottomrule
\end{tabular}}
\caption{Correlations between description length, similarity metrics, and each error type in annotated alt texts. Dashes indicate insignifigant correlations with p-values greater than 0.005.}
\label{table:correlations}
\end{table*}